\documentstyle[aps,manuscript,epsf]{revtex} % 12 pt
\def\baselinestretch{1.55}
\clearpage

\begin{document}

\input{psfig}
%\draft
%\begin{titlepage}
\begin{flushright}
CDF/PUB/BOTTOM/PUBLIC/4534\\
FERMILAB-Pub-98/167-E\\
%\today \\
June 14, 1998 \\
\end{flushright}
\vspace{0.20in}
\begin {center}
\begin {Large}
{\bf 
Improved measurement of the $B^-$ 
and ${\overline B}$$^0$ meson lifetimes
using semileptonic decays}
%\footnote{To be submitted to {\em Physical Review D}. }
\footnote{Submitted to {\em Physical Review D}. }
\end {Large}
\\
\end{center}
%%%%%%%%%%%%%%%%%%%%%%%%%%%%%%%%%%%%%%%%%%%%%%%%%%%%%%%%%%%%%%%%%%%%%%%%%%%
%%% CDF author list from carol, removed our own macros
%From:	FNALD::CAROL         6-MAY-1998 11:33:26.13
%To:	FNALD::UKEGAWA
%CC:	
%Subj:	CDF Default Author List!

%\font\eightit=cmti8
%\def\r#1{\ignorespaces $^{#1}$}
\hfilneg
\begin{sloppypar}
\noindent
F.~Abe,$^{17}$ H.~Akimoto,$^{39}$
A.~Akopian,$^{31}$ M.~G.~Albrow,$^{7}$ A.~Amadon,$^{5}$ S.~R.~Amendolia,$^{27}$ 
D.~Amidei,$^{20}$ J.~Antos,$^{33}$ S.~Aota,$^{37}$
G.~Apollinari,$^{31}$
% ...
\end{sloppypar}
\vskip .026in
\begin{center}
(CDF Collaboration)
\end{center}

\vskip .026in
\begin{center}
\begin{footnotesize}
% ...
$^{5}$  {\it University of Chicago, Chicago, Illinois 60637} \\
% ...
$^{7}$  {\it Fermi National Accelerator Laboratory, Batavia, Illinois 
60510} \\
% ...
$^{17}$ {\it National Laboratory for High Energy Physics (KEK), Tsukuba, 
Ibaraki 305, Japan} \\
% ...
$^{20}$ {\it University of Michigan, Ann Arbor, Michigan 48109} \\
% ...
$^{27}$ {\it Istituto Nazionale di Fisica Nucleare, University and Scuola
               Normale Superiore of Pisa, I-56100 Pisa, Italy} \\
% ...
$^{31}$ {\it Rockefeller University, New York, New York 10021} \\
% ...
$^{33}$ {\it Academia Sinica, Taipei, Taiwan 11530, Republic of China} \\
% ...
$^{37}$ {\it University of Tsukuba, Tsukuba, Ibaraki 315, Japan} \\
% ...
$^{39}$ {\it Waseda University, Tokyo 169, Japan} \\
% ...
\end{footnotesize}
\end{center}

\newpage
\vspace{0.25in}
\begin{abstract}
%\noindent
The lifetimes of the $B^-$ and $\overline{B}$$^0$ mesons are
measured using the partially reconstructed
semileptonic decays 
$\overline{B} \rightarrow D \ell^- {\overline \nu} X$,
where $D$ is either a $D^0$ or $D^{*+}$ meson.
The data were 
collected by the CDF detector at the Fermilab Tevatron collider
during 1992-1995 and correspond to about 110~pb$^{-1}$ of
$\bar{p} p$ collisions at $\sqrt{s} =$ 1.8 TeV.
We measure decay lengths %may 5 98   distributions 
and extract % may 5 98 find
the  lifetimes to be 
%$\tau(B^-) = 1.637 \pm 0.057 \pm 0.053$~ps 
%$\tau(B^-) = 1.637 \pm 0.057 \pm 0.051$~ps 
%$\tau(B^-) = 1.637 \pm 0.058 \pm 0.051$~ps 
$ \tau(B^-) = 1.637 \pm 0.058 \, ^{+\, 0.045} _{ -\, 0.043}$~ps 
and
%$ \tau({\overline B}$$^0) = 1.480 \pm 0.040 \pm 0.054$~ps,
%$ \tau({\overline B}$$^0) = 1.480 \pm 0.040 \pm 0.057$~ps,
%$ \tau({\overline B}$$^0) = 1.474 \pm 0.039 \pm 0.057$~ps,
$  \tau({\overline B}$$^0) = 1.474 \pm 0.039 
\, ^{+\, 0.052} _{ -\, 0.051}$~ps, 
and 
the  ratio of the lifetimes to be
$\tau(B^-) / \tau({\overline B}$$^0) = 
%1.106 \pm 0.055 \, ^{+\, 0.031} _{ -\, 0.028}$,
%1.106 \pm 0.055 \, ^{+\, 0.033} _{ -\, 0.029}$,
%1.110 \pm 0.056 \, ^{+\, 0.033} _{ -\, 0.029}$,
 1.110 \pm 0.056 \, ^{+\, 0.033} _{ -\, 0.030}$,  % may 6, 98
where the first uncertainties are statistical and the second are systematic.
%%%These results are in good agreement with previous measurements.
\end{abstract}
\vspace{1cm}
 \vspace{0.2in} \\
%\end{titlepage}
%

%%%% \newpage
%%%%%%%%%%%%%%%%%%%%%%%%%%%%%5 1) Introduction
\section{Introduction}
Measurements of the lifetimes of the individual $B$-hadron species
can probe their % $B$-hadron 
decay mechanism
beyond the simple spectator model decay picture.
In this model, all hadrons containing a heavy quark % ting $Q$ 
should have one identical lifetime, that of the quark. %ting $Q$.
%% hans This would be true ultimately for %abw in
%% the top quark, 
%%which decays even before physical hadrons are formed.
%% hans On the other hand, 
However, this picture does not hold in the case of
charm hadrons; the lifetimes of $D^+$ and $D^0$ mesons 
differ by a factor of 2.5. 
Possible causes of lifetime differences 
include contributions from non-spectator decays,
namely the annihilation and the $W$-exchange processes,
and so-called final-state  %abw final state 
Pauli interference effects.
Obviously, %abw, was no ","
these mechanisms play an important role 
in the decay of charm hadrons. 
However, they are expected to produce smaller lifetime differences
between %abw in 
the $B$ hadrons because of the larger mass of the $b$ quark.

In the past few years, 
the heavy quark expansion technique has been applied extensively
to the calculations of inclusive decay rates of heavy hadrons, 
both spectator and non-spectator decays. It provides quantitative
predictions for lifetime differences among the heavy hadrons.
It is generally believed that there should exist
a lifetime difference of order (5-10)\% 
between the $B^-$ and ${\overline B}$$^0$ mesons.
Bigi predicts~\cite{theory1} that the $B^-$ meson
lifetime should be longer than the ${\overline B}$$^0$ meson lifetime.
However, Neubert and Sachrajda~\cite{Neubert} state that 
the sign of the %abw
deviation from unity cannot be predicted reliably.
A much smaller difference, of order 1\%, is predicted
for 
the ${\overline B}$$^0$ and ${\overline B}$$^0_s$ meson lifetimes.

Several direct measurements of
$B^-$ and ${\overline B}$$^0$ meson lifetimes
have been performed recently 
by the $e^+e^-$ experiments~\cite{bu_bd_life}
and by CDF~\cite{semi_1a,cdf_psi_new}.
Indirect information has been obtained 
through the measurement of 
branching fractions~\cite{ARGUS}.  % add a reference Fumi
The precision of current measurements now
approaches the level where the predicted small differences
could be discerned, and improvements in these measurements
will provide a strong test of $B$-hadron decay mechanisms.

In this Article 
we report a measurement of the $B^-$ and ${\overline B}$$^0$  meson
%charged and neutral $B$ meson
lifetimes using partially reconstructed semileptonic decays.
The data used in this analysis
were collected 
in 1992-95
with the CDF detector
at the Fermilab Tevatron 
%${\overline p}p$ 
proton-antiproton
collider at a center-of-mass energy $\sqrt{s}=1.8$ TeV %abw ,
and correspond to an integrated luminosity of about 110 pb$^{-1}$.

In order to identify semileptonic decays of $B$ mesons, 
events with a lepton ($e^-$ or $\mu^-$, denoted by $\ell^-$)
associated with a $D^0$ or $D^{*+}$ meson 
are selected. 
(Throughout this Article a reference to a particular charge state
also implies its charge conjugate.)
The $\ell^- D^{0}$ candidates consist mostly of $B^-$ decays,
and the $\ell^- D^{*+}$ candidates 
consist mostly of ${\overline B}$$^0$ decays.
The $D^0$ mesons are %abw is 
reconstructed using the decay mode $D^0 \rightarrow
K^- \pi^+$.
The $D^{*+}$ decays are reconstructed 
using the decay mode $D^{*+} \rightarrow D^0 \pi^+$, followed by
$D^0 \rightarrow K^- \pi^+$, $K^- \pi^+ \pi^+ \pi^-$ or $K^- \pi^+ \pi^0$.
About 6000 such decays are reconstructed in the data sample.
%from the data sample corresponding to an integrated luminosity of
%19.3 pb$^{-1}$ recorded during 1992-1993. 
The decay length distributions are measured 
and                    %nov 30  with momentum estimates
the lifetimes are extracted
after correcting for the relative admixtures 
of %abw the 
$B^-$ and ${\overline B}$$^0$ mesons in the samples.
%%% ANDY +++
The results presented here %use the CDF data from 1992 to 1995, and 
supersede a previous CDF measurement~\cite{semi_1a},
since the part of the data sample used here is the %trischuk june 4
same as that of Ref.~\cite{semi_1a}.

% 2) CDF Detector
%
\section{CDF Detector and Trigger}
The CDF detector is described in detail elsewhere~\cite{CDF}. 
We describe here only the detector components 
most relevant to this analysis. 
Inside the 1.4 T solenoid
the silicon vertex detector (SVX) and the central tracking chamber (CTC)
provide the tracking and momentum analysis of charged particles.
The CTC is a cylindrical drift chamber containing 84 
measurement %readout 
layers.
% grouped into 9 alternating superlayers of axial and stereo wires. 
%The CTC
It 
covers the pseudorapidity interval $|\eta| < 1.1$, 
where $\eta=-\ln[\tan(\theta/2)]$~\cite{foot2}.
The SVX consists of four layers of silicon micro-strip detectors
located at radii between  3.0 and 7.9~cm from 
the interaction point 
and provides spatial measurements in the $r$-$\varphi$ plane
%%at radii between 2.9 and 7.9 cm,
with a resolution of 13 $\mu$m. %%%, giving 
It
gives
a track impact parameter resolution of about 
$(13 + 40/p_T)~\mu$m~\cite{svx}, 
where $p_{\rm T}$ is the transverse momentum of the track 
measured in GeV/$c$.
The geometric acceptance of the SVX is $\sim 60\%$, %abw 
as it extends to $\pm~25$ cm from the nominal
interaction point, %abw
whereas 
%%hans  the Tevatron beam has an 
the position of the primary interaction vertices
has an
rms width of $\sim 30$ cm along the beam ($z$) direction.
The transverse profile of the Tevatron beam
%%is circular and has an rms spread of $\sim 35$ $\mu$m. %Ia
is circular and has an rms spread of 
%franco $\sim 25$ $\mu$m.
$\sim 35$ $\mu$m for the data taking period in 1992-93
and $\sim 25$ $\mu$m in 1994-95.
%, while the longitudinal beam size is  $\sim 30$ cm.
The $p_T$ resolution of the CTC combined with the SVX is 
$\sigma(p_T)/p_T = [ (0.0066)^2 + (0.0009 \, p_T)^2 ]^{1/2} $.  
Electromagnetic (CEM) 
and hadronic (CHA) 
calorimeters
with projective tower geometry 
are located 
outside the solenoid
and
cover the pseudorapidity region $|\eta|<1.1$,
with a segmentation of $\Delta \varphi = 15^{\circ}$
and $\Delta \eta \simeq 0.11$.
A layer of proportional chambers (CES)  is embedded near shower
maximum in the CEM and 
provides a more precise measurement of electromagnetic 
shower profiles and an %abw
additional measurement of pulse height.
Also, %, abw
a layer of proportional chambers (CPR)
is installed between the solenoid and the CEM, and samples
%abw the 
electromagnetic showers at about one radiation length.
Two muon subsystems 
in the central rapidity region 
are used for muon identification.
The central muon chambers (CMU) 
are located just behind the CHA calorimeter,
and the central upgrade muon chambers (CMP) follow
an % Kroll 7 may 98
additional 60~cm of steel. 

Events containing semileptonic $B$ decays are collected using  
inclusive lepton triggers. 
CDF uses a three-level trigger system, where 
at the first two levels decisions are made with dedicated hardware.
The information available at this stage
includes energy 
deposits % plural kroll
in the CEM and CHA calorimeters,
high $p_T$ tracks found in CTC by a track processor,
and track segments found in the muon subsystems.
The $E_T$ %\; (\equiv E \sin \theta)$ 
threshold for the principal
single electron trigger is 9 (8) GeV
for the data taking period in 1992-93 (94-95),
where $E_T \equiv E\sin \theta$, %abw
and $E$ is the energy measured in the CEM.
In addition, %%a small leakage is required in the CHA,
%%as well as 
a track is required
in the CTC with $p_T > 7.5$ GeV/$c$
that points at the calorimeter tower in $\varphi$.
For the 1994-95 data taking period 
the CES was added to the trigger system~\cite{XCES}.
The electron trigger requires 
the presence of pulse height in the CES 
corresponding to an electromagnetic shower of 4 GeV or above.
Also, the $\varphi$ position of the shower  
is available with a segmentation of 
$\Delta \varphi = 2^{\circ}$, and the CTC track is required to
point at the shower.
The single muon trigger requires 
a track in the CTC,
corresponding to a particle with  $p_T > 7.5$ GeV/$c$,
and track segments in both the CMU and CMP systems
that match the CTC track within $7.5^{\circ}$ in $\varphi$.
%corresponding to 
%a particle with  $p_T > 7.5$ GeV/$c$.
%a $p_T$ threshold of 7.5 GeV/$c$.
%%%%% $p_T$ threshold is 7.5 GeV/$c$.
At the third level of the trigger, %hans May 05 essentially all the information
% is available, and 
the % kroll may 98
event selection is based on a
version of off-line reconstruction programs
optimized for speed.
The lepton selection criteria used in level 3 
are similar to those described  in the next Section.
% kroll some mods

%%%%%%%%%%%%%%%%%%%%%%%%%%%%%%%%%%%%%%%   B decay reconstruction
\section{Reconstruction of Semileptonic Decays of $B$ Mesons}
% Added for Ting, may 05, 98
The analysis starts with identification of leptons, $e^-$ or $\mu^-$.
If an event contains a good lepton candidate,
we look for the charm meson $D^{0}$ or $D^{*+}$
produced in the vicinity of the lepton candidate, 
to be consistent with the semileptonic decay signature
${\overline B} \rightarrow \ell^- \bar {\nu} D X$.
A proper correlation between the lepton charge
and the charm flavor, $\ell^-$ with $D$, not $\ell^+$ with $D$, 
is required.

\subsection {Lepton Identification}
The identification of electrons makes use of information 
from both calorimeters %(CEM, CHA) 
and tracking chambers.
To be specific we require the following:\\ %abw .
\begin{itemize}
\item Longitudinal profile consistent with an electron shower, i.e.
small leakage energy in the CHA.
\item 
Lateral shower profiles measured in the CEM~\cite{lateral1}
and the CES~\cite{lateral2} consistent with electron test beam data. 
\item Association of a high $p_T$ track with the calorimeter
shower based on position matching and energy-to-momentum ratio. 
\item Pulse heights in the CES and CPR consistent with an electron.
\end{itemize}
Photon conversion electrons, %abw %%Hans %% due to detector material, 
as well as %abw the 
Dalitz decays of $\pi^0$ mesons,
are removed by looking for oppositely charged tracks 
%kroll which 
that 
have small opening angles with the electron candidate.

%fumi feb 98 The criteria for the muon identification are
%described in Ref.~\cite{cdf_life}.
Muons are identified based on the geometrical matching between the
track segments in the muon chambers and an extrapolated CTC track.
%hans The specific criteria are described in Ref.~\cite{cdf_life}.
We compute the $\chi^2$ of the matching, where the uncertainty is 
dominated by multiple Coulomb scattering in the detector material. 
We require $\chi^2 < 9$ in the $r$-$\varphi$ view (CMU and CMP)
and $\chi^2 < 12$ in the $r$-$z$ view (CMU).

%fumi Feb 98, moved to below
%%In order to ensure accurate decay length measurement, lepton tracks
%%are required to be reconstructed in the SVX.

\subsection {Charm meson reconstruction}
\begin{figure}[p]
\vspace{1in}
\epsfysize=5.5in
%%\epsffile[ 100 144 522 648]{mass_1ab.ps}
\epsffile[ 100 144 522 648]{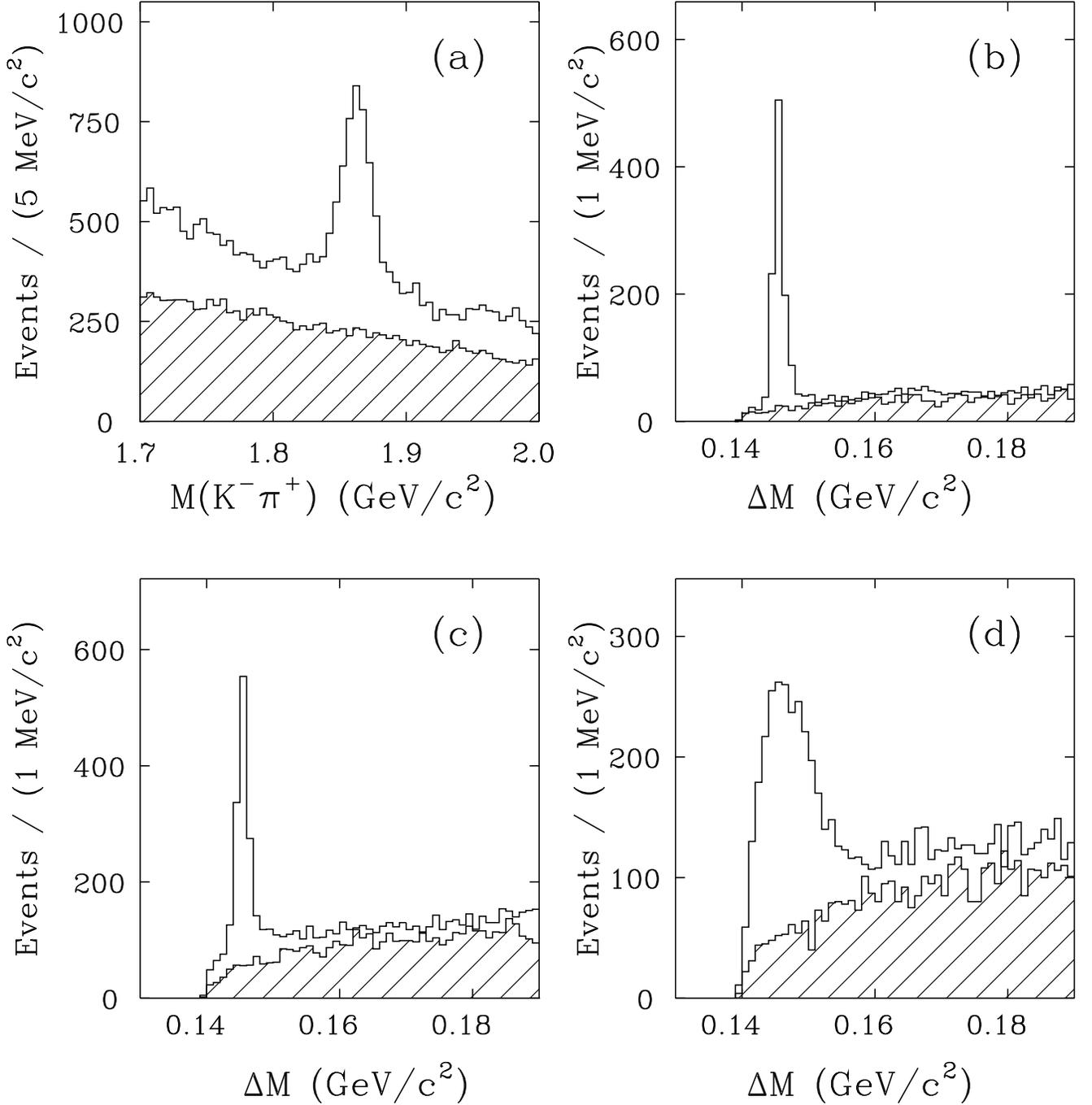}
\vspace{1in}
\caption{ Charm signals 
reconstructed in the vicinity of leptons $\ell^-$.
Four modes are shown: (a) $D^0 \rightarrow K^- \pi^+$ (non-$D^{*+}$), 
(b) $D^{*+} \rightarrow D^0 \pi^+, D^0 \rightarrow K^- \pi^+$,
(c) $D^{*+} \rightarrow D^0 \pi^+, D^0 \rightarrow K^- \pi^+ \pi^+ \pi^-$ and
(d) $D^{*+} \rightarrow D^0 \pi^+, D^0 \rightarrow K^- \pi^+ \pi^0$.
Plot (a) shows the $K^-\pi^+$ invariant mass spectra,
and (b-d) show the $\Delta m$ distributions.
Shaded histograms 
%in modes (b-d) 
show wrong sign %($D^0 \pi^-$) 
combinations, and in (a) they are scaled by 0.5 for display purposes.
}
\label {fig:signal}
\end {figure}

To identify the $\ell^-D^0$ candidates, we search for $D^0\rightarrow
K^-\pi^+$ decays near the leptons, removing events that are consistent
with the $D^{*+}\rightarrow D^0 \pi^+$ decay chain.
The $D^0 \rightarrow K^- \pi^+$ decay is
reconstructed as follows.
We first select oppositely charged pairs of particles using CTC tracks,
where
the kaon mass is assigned to the particle with the
same charge as the lepton
(called the ``right sign" combination),
as is the case in semileptonic $B$ decays.
The kaon (pion) candidate is then required to have
momentum above 1.5 (0.5) GeV/$c$,
and to be within a cone of radius $\Delta R = 0.6$ (0.7) around the lepton
in $\eta$-$\varphi$ space, where 
$\Delta R = \sqrt{ (\Delta \eta)^2 + (\Delta \varphi)^2 }$.
%fumi feb 98
To ensure accurate decay length measurement, 
each candidate track is required to be reconstructed in the SVX,  %the by hans
as well as the lepton track.
To reduce combinatorial background,
we require the decay vertex of the $D^0$ candidate
to be positively displaced 
along its flight direction  
in the transverse plane
with respect to 
the position of the  primary vertex.
The primary vertex is approximated 
by the beam position~\cite{cdf_psi_new,cdf_life}.
To remove events consistent with the decay chain $D^{*+}\rightarrow
D^0\pi^+$, we combine additional positive tracks with the $D^0$ candidate
and compute the mass difference ($\Delta m$) between the
$D^0\pi^+$ and the $D^0$, assigning the pion mass to the tracks.  %; %Ting
The $\Delta m$ resolution is measured to be 0.74 MeV/$c^2$.
%franco We remove events with $\Delta m$ values 
%franco between 0.142 and 0.148 GeV/$c^2$.
We remove the $D^0$ candidate if any track exists 
that gives a $\Delta m$ value between 0.142 and 0.148 GeV/$c^2$.
The resulting $K^- \pi^+$ invariant mass spectrum is shown 
in Fig.~\ref{fig:signal}(a).
%abw The signal is apparent.
We fit %the spectrum with 
a polynomial background and a Gaussian distribution %, abw
to the spectrum
and find a mass resolution of 
%%%may 98 11.7 MeV/$c^2$.  %% this is with 10 mev bin
11.3 MeV/$c^2$.  % 5 mev bin
Also shown by the shaded histogram is the mass spectrum 
for the ``wrong sign" ($K^+ \pi^-$ with $\ell^-$) combinations,
where no significant signal is observed.
We define the signal region to be in
the mass range from 1.84 to 1.88 GeV/$c^2$.
The total number of events in the signal region is 
5198,  %1233 
and the background fraction 
is estimated from the fit to be 
$0.53 \pm 0.02$.
%$0.53 \pm 0.03$.

To identify  $\ell^-D^{*+}$ candidates, 
we search for $D^{*+} \rightarrow D^0 \pi^+$ decays
using 
two fully reconstructed $D^0$ decay modes,
$D^0 \rightarrow K^- \pi^+$ and
$D^0 \rightarrow K^- \pi^+ \pi^+ \pi^-$, and
one partially reconstructed mode,
$D^0 \rightarrow K^- \pi^+ \pi^0$.
%franco Similar kinematic cuts are applied to $D^0$ daughter candidates.
For the  $D^0 \rightarrow K^- \pi^+$ and $K^- \pi^+ \pi^0$ modes,
we apply the same momentum and cone requirements
to the kaon and pion candidates
as in the ${\overline B} \rightarrow \ell^- D^0 X$ reconstruction.
For the $D^0 \rightarrow K^- \pi^+ \pi^+ \pi^-$ mode,
the kaon (pion) candidate is required to have
momentum above 1.2 (0.5) GeV/$c$,
and to be within a cone of radius 0.65 (1.0) around the lepton candidate.
Also, %, abw
we require % harvard may 05 % positive decay lengths 
the decay vertex of the $D^0$ candidate
to be positively displaced with respect to the primary vertex
in the %% last two modes.
$D^0 \rightarrow K^- \pi^+ \pi^+ \pi^-$ and
$K^- \pi^+ \pi^0$ modes.
%%% They are listed in Table~\ref{tb:charm_cuts}.
For the fully reconstructed modes,
the $D^0$ candidate has to be in the mass ranges
1.83 to 1.90 GeV/$c^2$ and 1.84 to 1.88 GeV/$c^2$, respectively.
For the partially reconstructed mode, 
we require the mass of a $K^- \pi^+$ pair
to be between 1.5 and 1.7 GeV/$c^2$;
we do not reconstruct the $\pi^0$ and
in the subsequent analysis
treat the $K^-\pi^+$ pair as if it were a $D^0$.
For each mode, we reconstruct the $D^{*+}$ meson
by combining an additional track,
assumed to have the pion mass, with the $D^0$ candidate,
and computing the mass difference, $\Delta m$, between the
$D^0 \pi^+$ and $D^0$ candidates.
%For the last mode, the
%mass difference is defined as 
%       $\Delta m \equiv m(K^-\pi^+ \pi) -m(K^- \pi^+)$.
%apr 96 Here $D^0 \pi^+$ is the ``right sign" combination, and
%apr 96 $D^0 \pi^-$ is the ``wrong sign".
Figures~\ref{fig:signal}(b)-(d) show the $\Delta m$ distributions. 
%abw , where signals are apparent.
In Fig.~\ref{fig:signal}(d) 
the peak is broadened because of the missing $\pi^0$ meson.
Also shown by the shaded histograms 
are the spectra from the ``wrong sign" 
low-energy pion ($D^0 \pi^-$) combinations.
%% where no significant signals are observed.
We define the signal region as follows.
The two fully reconstructed modes
use the $\Delta m$ range 0.144 to 0.147 GeV/$c^2$, and
the $K^- \pi^+ \pi^0$ mode uses the 
range $\Delta m <0.155$ GeV/$c^2$.
The numbers of events in the signal regions are
%%200, 332 and 704, %in the three $D^{*+}$ modes,
935, 1166, and 2858, %abw .  %2820.
respectively.

We estimate the numbers of combinatorial background events 
by using the shapes of the $\Delta m$ spectra of 
the wrong sign ($D^0 \pi^-$) combinations
and normalizing them 
to the number of events %% hans
in the $\Delta m$ sideband.
The estimated background fractions    % in the signal regions are 
are
$0.09 \pm 0.01$, $0.18 \pm 0.02$ and $0.37 \pm 0.02$, respectively.
They are summarized in Table~\ref{tb:signal}.

It is possible that 
real $D^0$ or $D^{*+}$ mesons 
are accompanied
by %abw with 
%may 5 98 a misidentified hadron $h^-$, %% (fake lepton), %% lepton 
a hadron $h^-$ that was misidentified as a lepton,
%% or with a real lepton that is unrelated with the charm meson
and such events can be included in the above samples.
The hadrons can be either the decay products of the same $B$ hadron
that produced the charm meson
or the primary particles produced in 
${\bar p} p \rightarrow b{\bar b}X$ 
and   
${\bar p} p \rightarrow c{\bar c}X$ 
events.
We investigate this possibility
by studying %looking at
the wrong sign combinations,
$\ell^+$ with $D^0$ or $D^{*+}$, which 
cannot originate from $B$ meson decays.
We see no evidence for signal in these combinations. %abw those pairs, 
%% nov 97   obtain upper limits on the fractions of such events
%% nov 97   to be ??.??.
%%% nov 97 +++
Based on this study % kroll
we estimate 
the contribution of the $D^{(*)}h^-$ pairs
to our signal to be
%$0.5 \pm 1.1$\% 
%$0.6 \pm 0.9$\%    % from l+ D*+
%may 5, 98$(1.2 \pm 2.4)$\%,
$(1.2 \, ^{+\,2.4} _{-\,1.2})$\%,
% from l+ D0. Uses 1:1 D0 h- vs D0 h+.
where possible charge correlations between the charm meson 
and the hadrons are considered.
%%% of the $\ell^- D^{(*)}$ signals.
We ignore this background, and treat it as a systematic uncertainty.
%%% Novv 97 ---

\section {Decay length measurement and momentum estimate}
\begin{figure}[p]
\vspace{1.5in}
\epsfysize=6.0in
\epsffile[ 50 144 522 648]{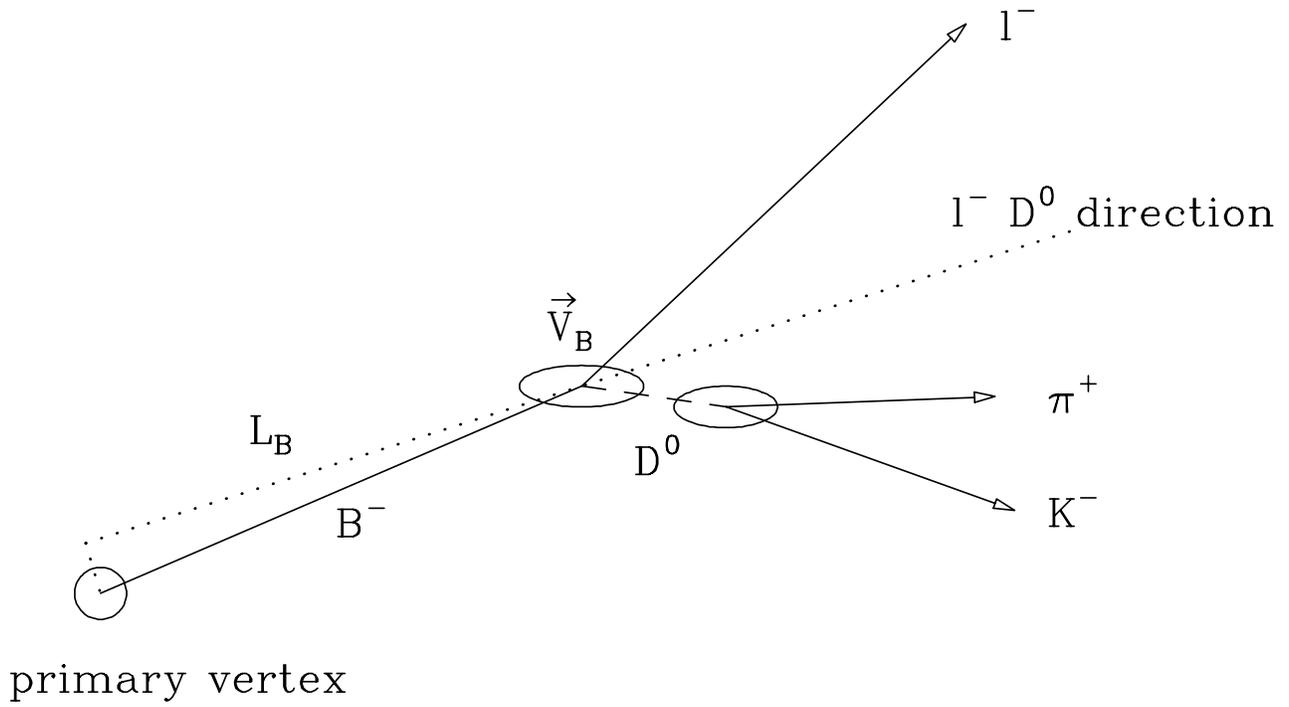}
\vspace{1in}
\caption
{Schematic representation of the decay
$B^- \rightarrow \ell^- {\bar \nu} D^0 X$,  %Andy
$D^0 \rightarrow K^- \pi^+$.
}
\label {fig:schematics}
\end {figure}

A schematic representation of the $B$ meson semileptonic decay 
topology is illustrated in Fig.~\ref{fig:schematics}.
The $B$ meson decay vertex ${\vec V}_B$
is obtained by
intersecting the trajectory of the lepton track with the flight path
of the $D^0$ candidate. %%%%~\cite{satellite}.
The $B$ decay length $L_B$ is defined as the displacement 
%%%in the transverse plane  %%% Hans W
of ${\vec V}_B$ from the primary vertex ${\vec V}_P$,
measured in the plane perpendicular to the beam axis, and 
projected onto the transverse momentum vector
of the lepton-$D^0$ system:
\[ L_B \equiv 
\frac { ( {\vec V}_B - {\vec V}_P ) \cdot {\vec p}_T^{\, \ell^- D^0} }
      { p_T^{\ell^-D^0} }.
\]

To measure a proper decay length %hans time 
of a $B$ meson decay, 
we need to know the momentum of the $B$ meson.
In semileptonic decays, 
the $B$ meson momentum cannot be measured precisely
because of the missing neutrino.
We use $p_T^{\ell^-D^0}$ to estimate the $B$ momentum for each event,
which results in a corrected decay length defined as
\[ x = L_B \, m_B /p_T^{\ell^- D^0}.
\]
We call it the `pseudo-proper decay length'.
The residual correction between
$p_T^{\ell^- D^0}$ and $p_T^B$ 
is performed %HArvard May 05 % statistically
during lifetime fits
we shall describe later.  %% added, kroll may 98

A typical resolution on this decay length $x$ due to vertex determination
is 50 $\mu$m,
including the contribution from the finite size of the primary vertex.
%kroll This resolution cannot be ignored compared with the $B$ meson
%lifetime,
%therefore we use a probability density distribution which
%includes the smearing. 
For subsequent lifetime measurements, we use only those events
%which have 
%hans may 05 98 reconstructed decay lengths $x$ 
%in the range between $-0.15$ cm and 0.3 cm, 
in which 
the resolutions on reconstructed decay lengths $x$ are 
smaller than 0.05 cm. 
Also we require the proper decay length of 
the $D^0$ meson, measured from the $B$ meson decay vertex to
the $D^0$ decay vertex, 
to be in the range from $-0.1$~cm to 0.1~cm
with its uncertainty smaller than 0.05~cm.
These cuts reject %reduce
poorly measured decays and 
reduce % added may 5
random track combinations.
In addition, we limit ourselves to events with
reconstructed decay lengths $x$ 
in the range between $-0.15$ cm and 0.3 cm.
These cuts have been applied already 
for the charm signals shown in Fig.~\ref{fig:signal}.

\begin{figure}[p]
\vspace{1.5in}
\epsfysize=7.0in
%%% JAN 98 \epsffile[ 100 144 522 648]{LIFE_ED_K_D0.ps}
%%% \epsffile[ 100 144 522 648]{RAT_ED0_P_ALL.ps}
\epsffile[ 100 144 522 648]{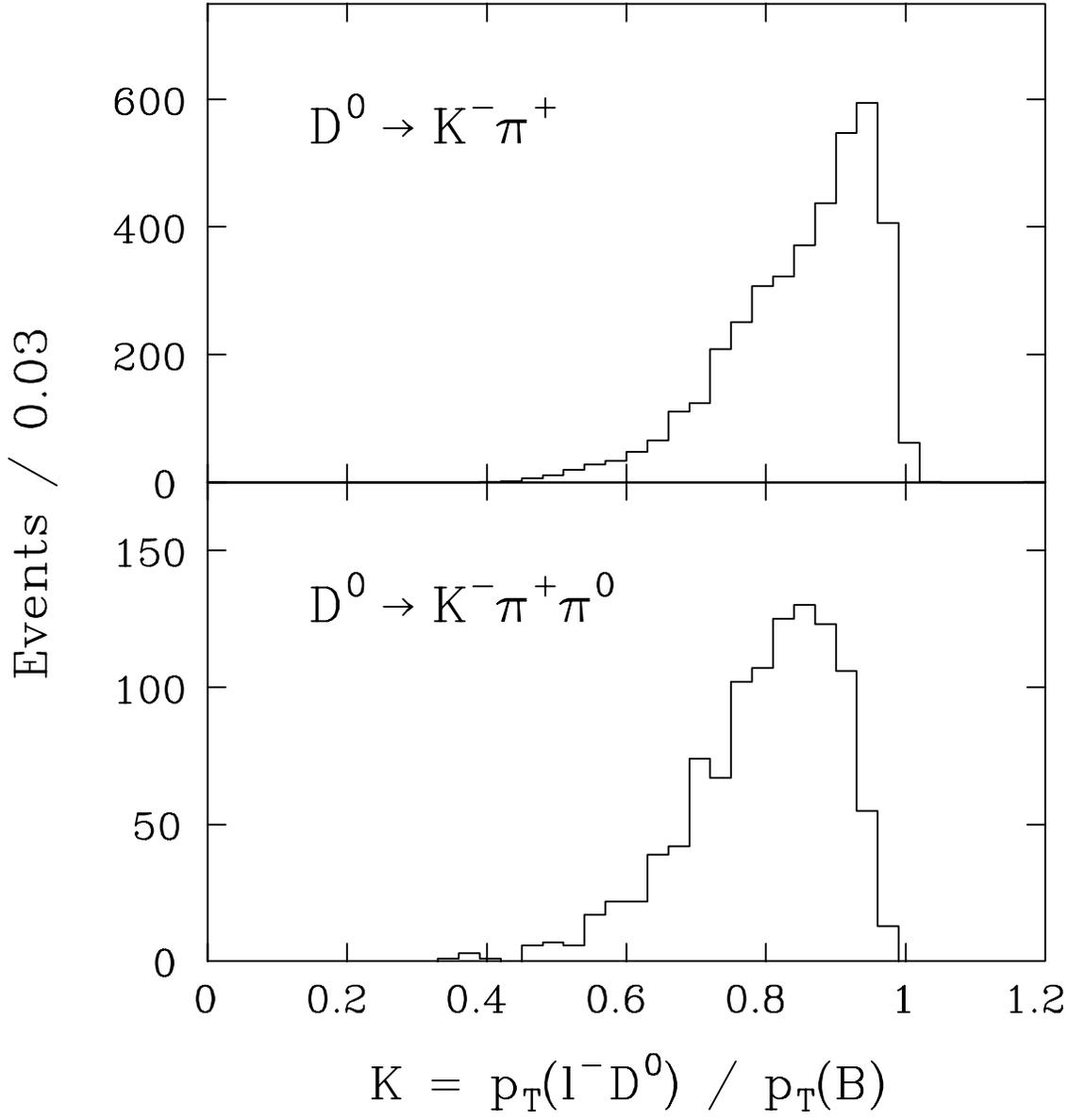}
\vspace{1in}
\caption
{ Distribution
of the momentum ratio $K$ (see text)
for
${\overline B} \rightarrow \ell^- {\bar \nu} D^0 X$, 
followed by
$D^0 \rightarrow K^- \pi^+$ and
$D^0 \rightarrow K^- \pi^+ \pi^0$ 
decays
obtained from a Monte Carlo calculation.
}
%for electrons above (a) 7 GeV/c and (b) 12 GeV/c.}
\label {fig:K}
\end {figure}

%%%%%% fitting procedure
As mentioned above, we have used the momentum of the 
lepton-$D^0$ system, $p_T^{\ell^- D^0}$, to calculate
the pseudo-proper decay length. However, we still need to 
account for the missing momentum to measure $B$ meson lifetimes.
We define the ratio $K$ of the observed momentum to the true momentum
as
\[
	K = p_T^{\ell^- D^0}/p_T^{B}.
\]
The $K$ distribution is obtained from a Monte Carlo calculation.
The ISAJET event generator~\cite{ISAJET} 
is used for the production of the $b$ quark,
where the shape of the $p_T$ spectrum is modified
slightly to match the QCD calculation in the next-to-leading order~\cite{NDE}.
The fragmentation model by Peterson and others~\cite{Peterson}
is used.
%%with the parameter $\epsilon = 0.006$~\cite{fragmentation} is used.
The CLEO event generator~\cite{QQ} is used to describe $B$ meson decays. 
In particular, the semileptonic
decays adopt the model by Isgur and others (ISGW)~\cite{ISGW}.
A typical $K$ distribution thus obtained 
has an average value of 0.85 with an rms width of 0.11,
and is approximately independent of
the $p_T^{\ell^- D^0}$
in the range of interest, 
which is typically 15 to 25 GeV/$c$.
It is also independent of 
the $D^0$ decay mode except for the
partially reconstructed mode $D^0 \rightarrow K^- \pi^+ \pi^0$,
%%% which has a slightly softer distribution (average of 0.80) %% Hans W
which has a slightly lower mean value (about 0.80) 
because of the missing $\pi^0$ particle.
%%An example of such a distribution is shown in Fig.~\ref{fig:K}.
%kroll An example of 
Two %kroll
$K$ distributions 
are 
shown in Fig.~\ref{fig:K}.

The lifetime is determined from a maximum likelihood fit to 
the  %% Hans W.
observed pseudo-proper decay length distributions.
The likelihood  for 
%kroll  a 
the 
signal sample is given by
\begin{displaymath}
{\cal L}_{\rm SIG}  =  \prod_i [ (1-f_{\rm BG}) {\cal F}_{\rm SIG}(x_i)
 + f_{\rm BG} {\cal F}_{\rm BG} (x_i) ],
\end{displaymath}
where 
$x_i$ is the pseudo-proper decay length measured for event $i$,
and the product is taken over observed events in the sample.
The first term in the likelihood function
represents %%%the probability density function for 
a $B$ decay signal event, while
the second term accounts for combinatorial background events
whose fraction in the sample is $f_{\rm BG}$.
The signal probability density function  ${\cal F}_{\rm SIG}(x)$
consists of an exponential decay function 
$\frac{K}{c\tau} \exp ( - \frac {K x }{ c\tau} )$
defined for % only 
positive decay lengths,
smeared with a normalized $K$ distribution $D(K)$
and a Gaussian distribution with width $s \sigma_i$:
\[
{\cal F}_{\rm SIG}(x) = \int dK \, D(K)
\, 
\left[
\theta(x) \,
\frac{K} { c \tau} 
\exp \left( - \frac{Kx} { c \tau} \right)
\otimes G(x)
\right]
,
\]
where 
$\tau$ is the $B$ meson lifetime, $c$ is the speed of light,
$\theta(x)$ is the step function defined as 
$\theta(x)=1$ for $x \geq 0$ and 
$\theta(x)=0$ for $x  < 0$,
and the symbol ``$\otimes$" denotes a convolution.
$G(x)$ is the Gaussian distribution given by
\[
	G(x) = \frac{ 1} { s \sigma_i \sqrt{ 2 \pi } }
	  \, \exp \left( - \frac { x^2 } { 2 s^2 \sigma_i^2 } \right),
\]
and $\sigma_i$ is the estimated resolution on $x_i$.
The scale factor $s$ 
%fumi Feb 98
is introduced as a fit %kroll fitting 
parameter and
accounts for a possible incompleteness
of our estimate of the decay length resolution.
%fumi Feb 98It is very hard to perform 
The integration over the momentum ratio $K$
%%Therefore we approximate it 
is approximated by %%with 
a finite sum
\[
\int dK D(K) \rightarrow \sum_j D(K_j) \Delta K,
\]
where the sum is taken over bin $j$
of a histogrammed distribution $D(K_j)$ with bin width $\Delta K$.

	The pseudo-proper decay length 
distribution
of combinatorial background events, ${\cal F}_{\rm BG}(x)$, 
is measured %modeled %described 
using mass sideband events, 
assuming that
they represent the combinatorial background events under signal mass peaks.
The functional form of the distribution is
parameterized empirically
by a sum of a Gaussian distribution centered at zero,
and positive and negative exponential tails %.
smeared with a Gaussian distribution:
\begin {eqnarray*}
{\cal F} _{\rm BG} ( x )  & = & 
 (1-f_- - f_+) \,G( x ) \nonumber \\
& + & ( f_+ / \lambda_+ ) \, \theta(x) \exp ( - x / \lambda_+ ) 
\otimes G ( x ) 
\nonumber \\
&  + & ( f_- / \lambda_- ) \, \theta(-x) \exp ( + x / \lambda_- ) 
\otimes G ( x ).
%\label{eq:control}
\end {eqnarray*}
%hans The central Gaussian accounts for a zero-lifetime component, and
%hans the exponential tails account for the combinatorial background 
%hans due to $B$-hadron  decays.
The shape of the background function 
(parameters $f_{\pm}$ and $\lambda_{\pm}$)
and the 
resolution % added may 5
scale factor $s$,
as well as the signal lifetime $c \tau$,  % c added may 5
are determined from a simultaneous fit 
to a signal sample and a background sample.
We use the combined likelihood ${\cal L}$ defined as
${\cal L} =  {\cal L}_{\rm SIG} \; {\cal L}_{\rm BG} $,
where ${\cal L}_{\rm BG}  =  \prod_k {\cal F}_{\rm BG} (x_k)$
and  the product 
is taken over event $k$ in the background sample.
The amount of combinatorial background $f_{\rm BG}$ is also
a  parameter in the simultaneous fit.
This parameter is constrained by adding a term
$\frac{1}{2} \chi^2= \frac{1}{2} 
(f_{\rm BG}- \langle f_{\rm BG} \rangle)^2 / \sigma_{\rm BG}^2$  to
the negative log-likelihood $- \ell = - \ln {\cal L}$.
The average background fraction $\langle f_{\rm BG} \rangle$
and its uncertainty $\sigma_{\rm BG}$
are
estimated from %a fit to 
the signal mass distributions (Table~\ref{tb:signal}).

The background sample for  the $\ell^-D^0$ mode
is  formed from the $D^0$ sidebands, 
%defined by the mass ranges 1.72 to 1.80 and 1.92 to 2.00 GeV/$c^2$.
defined  by the mass ranges 1.74 to 1.79 and 1.94 to 1.99 GeV/$c^2$.
For the $\ell^- D^{*+}$ samples
we use $\Delta m$ sidebands:
we use the right sign ($D^0 \pi^+$) sideband
$0.15 < \Delta m < 0.19$ GeV/$c^2$
for the two fully reconstructed $D^0$ modes,
and $0.16 < \Delta m < 0.19$ GeV/$c^2$ 
for the $D^0 \rightarrow K^- \pi^+ \pi^0$ mode.
%franco We also use the wrong sign pion combinations 
%franco   in the range $\Delta m < 0.19$ GeV/$c^2$ for all three $D^0$ 
%franco   decay modes.
The background samples are summarized in Table~\ref{tb:control}.

\begin{figure}[p]
\vspace{1.5in}
\epsfysize=7.0in
%%\epsffile[ 50 144 542 648]{fig2n.ps}
%%\epsffile[ 50 144 522 648]{life_1ab_bg.ps}
\epsffile[ 50 144 522 648]{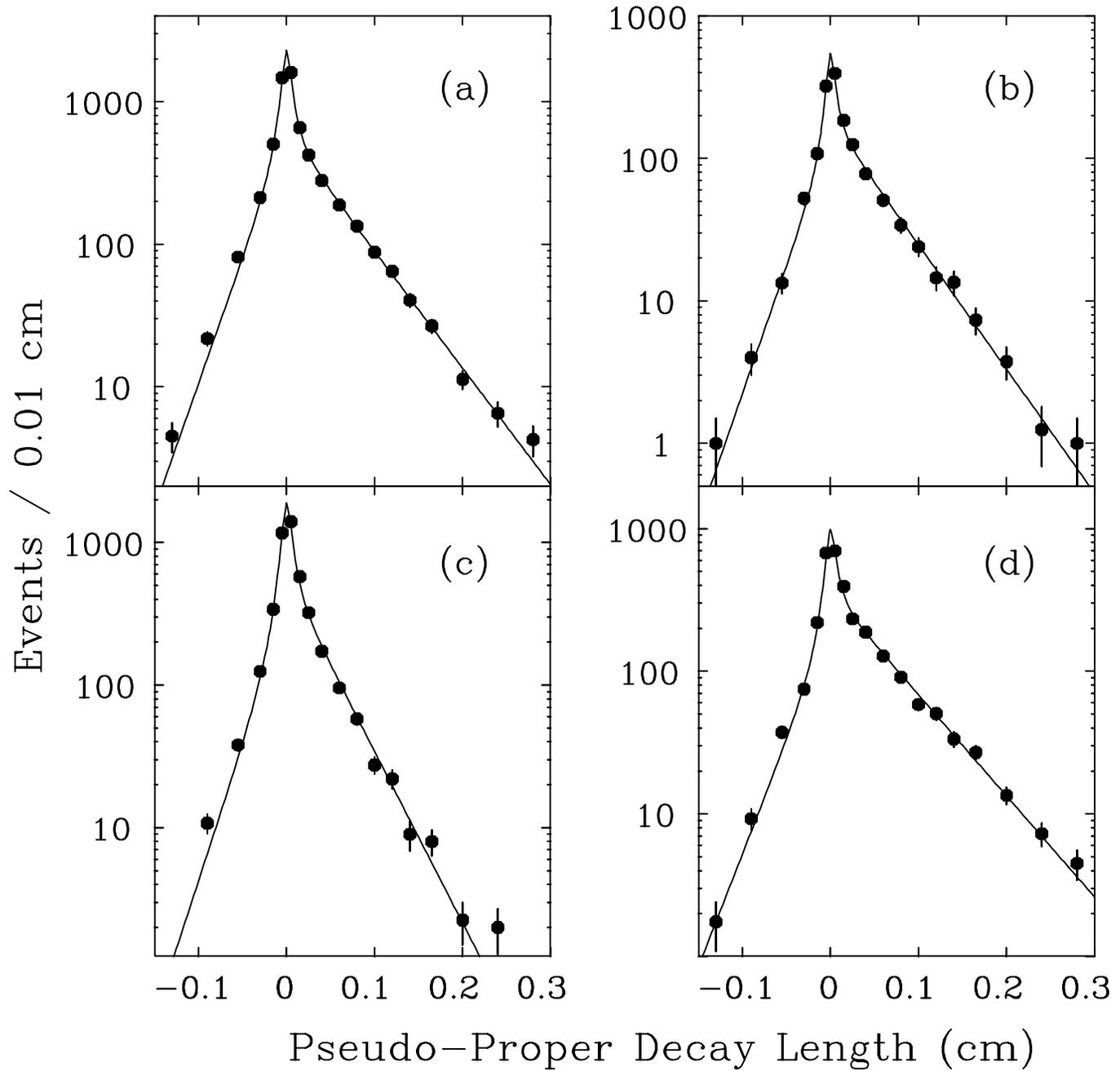}  % mar 98
\vspace{1in}
\caption
{Distributions of pseudo-proper decay lengths 
for lepton-$D$ background samples (points).
Also shown by the curve is the result of lifetime fits.
Four decay modes are shown:
(a) ${\overline B} \rightarrow \ell^- {\bar \nu} D^0 X$, 
    $D^0 \rightarrow K^- \pi^+$ (non-$D^{*+}$), 
and ${\overline B} \rightarrow \ell^- {\bar \nu} D^{*+} X$, 
$D^{*+} \rightarrow D^0 \pi^+$, followed by 
	(b) $D^0 \rightarrow K^- \pi^+$,
	(c) $D^0 \rightarrow K^- \pi^+ \pi^+ \pi^-$ and
	(d) $D^0 \rightarrow K^- \pi^+ \pi^0$.
}
\label {fig:bg}
\end {figure}

\begin{figure}[p]
\vspace{1.5in}
\epsfysize=7.0in
%\epsffile[ 50 144 522 648]{life_1ab.ps}
\epsffile[ 50 144 522 648]{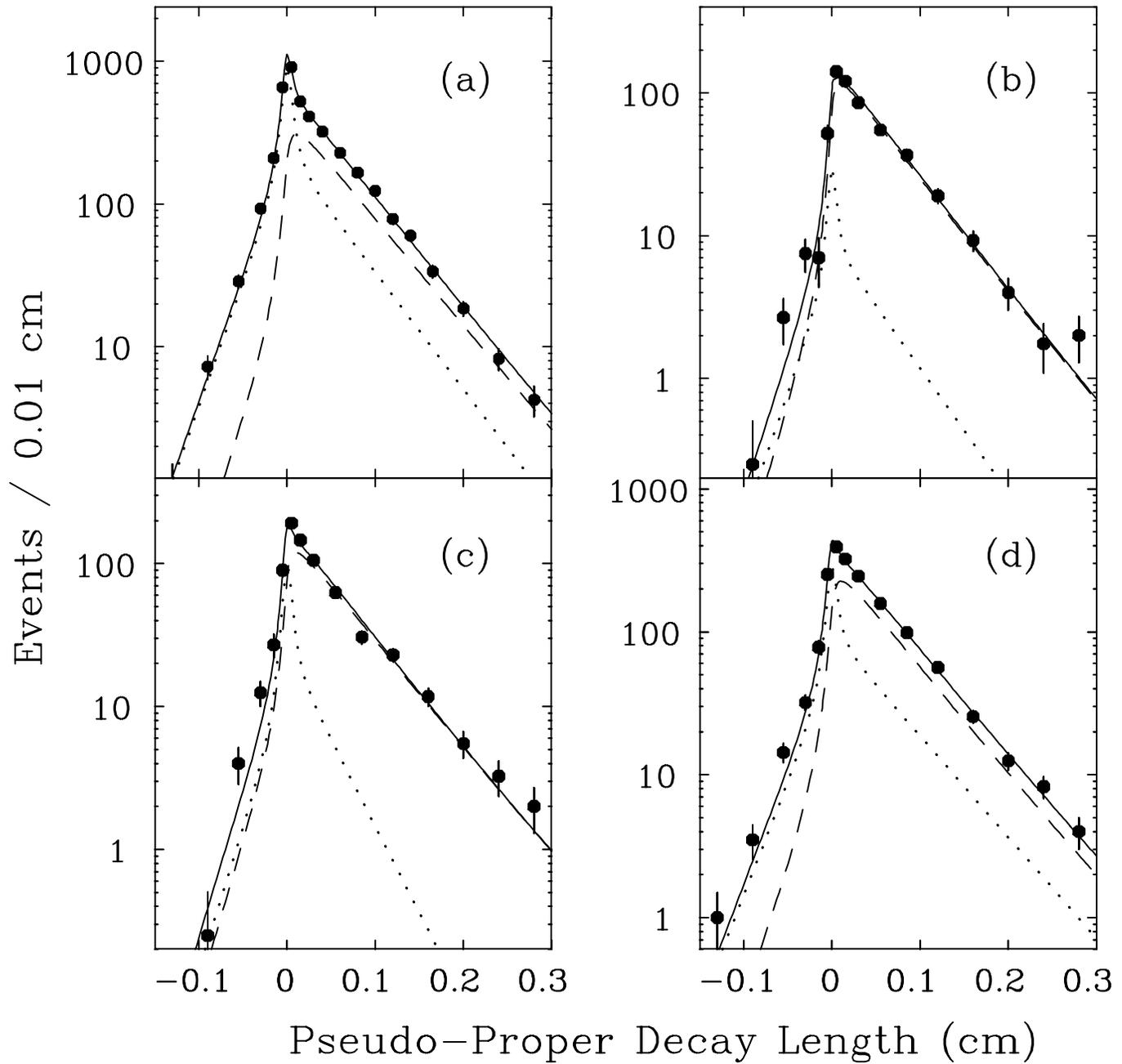}

\vspace{1in}
\caption
{Distributions of pseudo-proper decay lengths 
for lepton-$D$ signal samples (points).
Also shown are the result of lifetime fits,
signal (dashed curve) and background (dotted curve)
contributions, and the sum of the two (solid curve).
The four decay modes (a-d) are the same as in Fig.~{\protect \ref{fig:bg}}.
}
\label {fig:sig}
\end {figure}

%%% July 1997 addition
The pseudo-proper decay length distributions of the background samples
are shown in Fig.~\ref{fig:bg}, 
together with fit results.
The background parameter values and the resolution scale $s$ 
determined from the fit are listed in
Table~\ref{tb:shapes}.
The corresponding 
decay length  % added may 5
distributions of the signal samples
are shown in Fig.~\ref{fig:sig}.
We find the lifetimes to be
$c\tau(B)= 489 \pm 15, 
%461 \pm 18, 472 \pm 19$ and $453 \pm 15$ $\mu$m,
462 \pm 18, 472 \pm 19$ and $449 \pm 14$ $\mu$m
for the four modes,  % added may 5, 98
where uncertainties are statistical only.

\begin{figure}[p]
\vspace{1in}
\epsfysize=7.0in
%\epsffile[ 50 144 522 648]{life_d0_1ab.ps}
%\epsffile[ 50 144 522 648]{life_d0_1ab_sig_2exp.ps}
% MAR 14, 98  USE RS SIDEBAND ONLY AS BG
\epsffile[ 50 144 522 648]{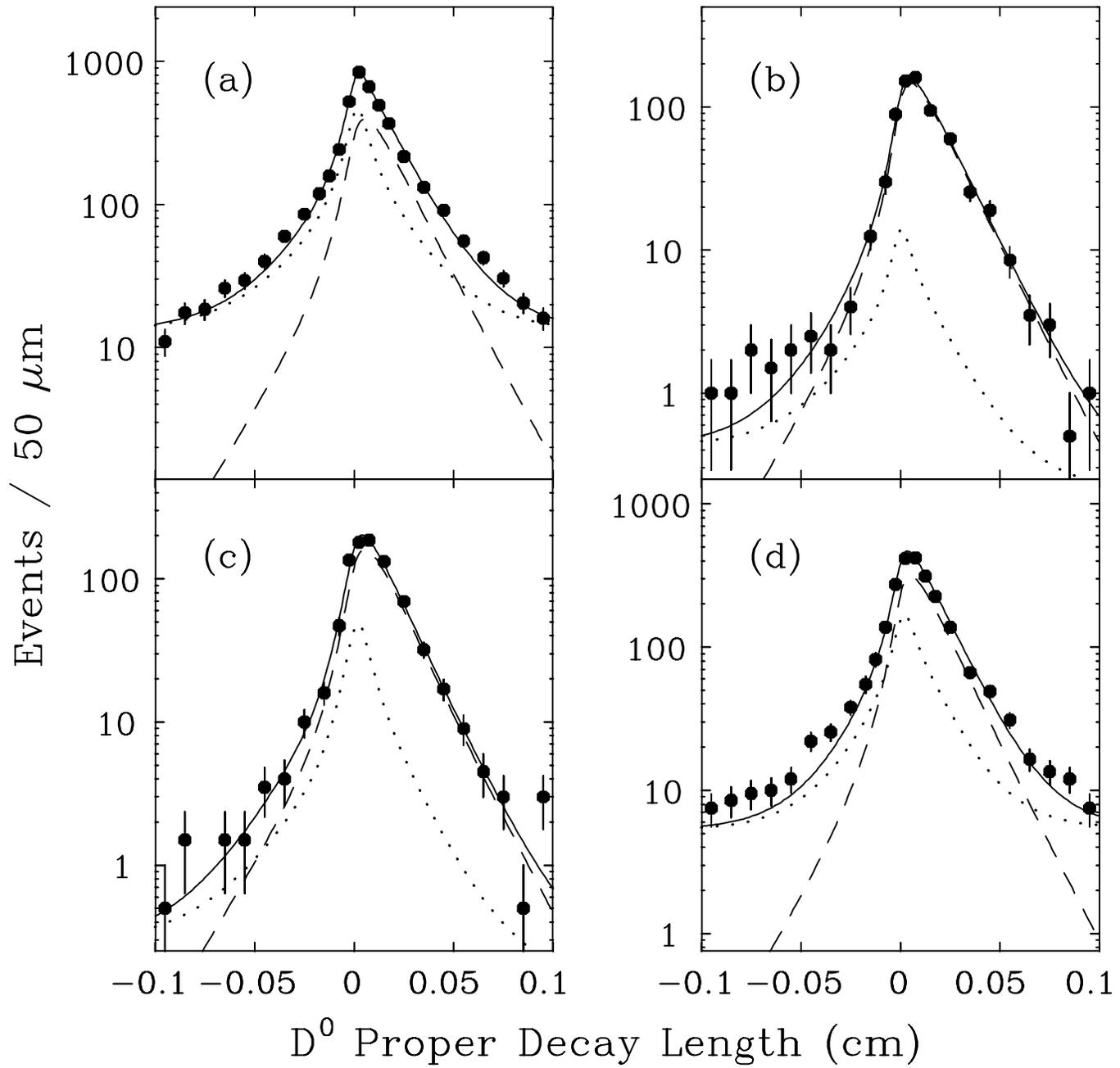}

\vspace{1in}
\caption
{Distributions of the $D^0$ proper decay lengths 
measured with respect to the $B$ meson decay vertex
%for individual lepton-$D$ signal samples 
(points).
Also shown are the result of lifetime fits,
signal (dashed curve) and background (dotted curve)
contributions, and the sum of the two (solid curve).
The four decay modes (a-d) are the same as in Fig.~{\protect \ref{fig:bg}}.
}
\label {fig:D0}
\end {figure}

As a check of the procedure,
we measure the $D^0$ lifetime 
using the  proper decay length 
measured from the secondary vertex ${\vec V}_B$ to the $D^0$ decay vertex.
The proper decay length distributions are shown in
Fig.~\ref{fig:D0}, together with fit results.
The lifetime numbers are summarized in Table~\ref{tb:lifetimes}.
The result
%%%%$133 \pm 12$ $\mu$m averaged over the samples,
is 
%$c\tau(D^0)= 144 \pm 12$, $132 \pm 13$,
%$132 \pm 12$ and $124 \pm 10$ $\mu$m 
%for the four modes,
%where the quoted uncertainties are statistical.
%They are 
in reasonably good agreement with 
the world average value of $124.4 \pm 1.2$ $\mu$m~\cite{PDG}.

\section { $B^-$ and ${\overline B}$$^0$ meson lifetimes}

In order to extract the $B^-$ and ${\overline B}$$^0$ meson lifetimes,
we must take into account the fact that the 
$\ell^- D^0$ and $\ell^- D^{*+}$  samples
are admixtures of the two $B$ meson decays.
The semileptonic decays  
can be expressed as
${\overline B} \rightarrow \ell^- {\overline \nu} {\bf D}$, where
${\bf D}$ is a charm system whose charge %($0$ or $+1$) 
is correlated with the $B$ meson charge. %($-1$ or $0$).
%%% If only the two lowest lying charm mesons, %% Hans W
If only the two lowest mass charm states, 
pseudoscalar ($D$) and vector ($D^*$),
are produced,
the $\ell^- D^{*+}$ combination
can arise only from the ${\overline B}$$^0$ meson decay.
Similarly, %abw  
the $\ell^- D^0$ combination comes only from $B^-$ meson decays,
provided that the $D^0$ from the $D^{*+}$ decay is excluded.
However, it is %abw has been 
known that 
%% the above two lowest lying states %% Hans W
the above two states
do not saturate the total semileptonic decay rates.
%And %abw 
%% It is generally believed %Trischuk June 98
All data indicate
that higher mass charm mesons, $D^{**}$ states,
as well as non-resonant $D^{(*)} \pi$ pairs,
are responsible for the rest of the
semileptonic decays. 
We do not distinguish resonant and non-resonant components,
and refer to both of them as $D^{**}$.

These $D^{**}$ meson decays can dilute the charge correlation
between the 
observed  %% added may 5, 98
final states and the parent $B$ meson.
For example, 
%%the $D^{**+}$ meson decays to both 
%%$D^{(*)+} \pi^0$ and $D^{(*)0} \pi^+$, %%and 
the $D^{**0}$  meson 
decays to %both %abw 
$D^{(*)+} \pi^-$ as well as %and 
$D^{(*)0} \pi^0$ final states, %abw
resulting in misidentification of
$B^-$ meson decays as
${\overline B}$$^0 \rightarrow D^{*+} \ell^- {\bar \nu} X$.
%abw Therefore the observed $\ell^- D^0$ and $\ell^- D^{*+}$ combinations
%abw are no longer pure samples of 
%abw $B^-$ and ${\overline B}$$^0$ meson decays.
Nevertheless, %abw they 
$\ell^- D^0$ and $\ell^- D^{*+}$ combinations
are dominated by %%%nearly independent samples %coming from in terms of 
%charged and neutral $B$ 
$B^-$ and ${\overline B}$$^0$ 
meson decays, respectively. %trischuk, 
As described below,
the contamination of the wrong $B$ meson species
is only at 10-15\% level.
% and 
This 
enable us to extract the
two $B$ meson lifetimes.

\subsection {Sample composition}
We estimate the fraction of $B^-$ decays $g^-$  %kroll
in the $\ell^- D^0$ and $\ell^- D^{*+}$  samples as follows.
The production rates of charged and neutral $B$ mesons
and their semileptonic decay widths are assumed to be equal. 
We also assume the $D^{**}$ meson 
decays exclusively to $D^{(*)}\pi$
via the strong interaction, thereby allowing us to 
determine the branching fractions, 
e.g.~$D^{(*)+} \pi^0$ vs $D^{(*)0} \pi^+$, 
using isospin symmetry. 
We consider four factors affecting the composition.
%abw Firstly, 
First, 
the composition depends on
the fraction ($f^{**}$) of the $D^{**}$ mesons 
produced in semileptonic $B$ decays,
\[
f^{**} = 
\frac { {\cal B} ({\overline B} \rightarrow \ell^- {\bar \nu} D^{**}) }
%{ {\cal B} ({\overline B} \rightarrow \ell^- {\bar \nu} D ) 
%+ {\cal B} ({\overline B} \rightarrow \ell^- {\bar \nu} D^{*} ) 
%+ {\cal B} ({\overline B} \rightarrow \ell^- {\bar \nu} D^{**} ) }. 
 { {\cal B} ({\overline B} \rightarrow \ell^- {\bar \nu} D X ) }
= 1 - \frac
 { {\cal B} ({\overline B} \rightarrow \ell^- {\bar \nu} D ) 
 + {\cal B} ({\overline B} \rightarrow \ell^- {\bar \nu} D^{*} ) }
 { {\cal B} ({\overline B} \rightarrow \ell^- {\bar \nu} D X )   }.
\]
The CLEO experiment measures the fraction of exclusive
decays to the two lowest mass states to be 
$0.64 \pm 0.10 \pm 0.06$~\cite{CLEO}.
Thus, %abw
we find that $f^{**} = 0.36 \pm 0.12$.
A few experiments have recently observed
some $D^{**}$ modes~\cite{ddst}, but the sum of exclusive modes
still does %abw do 
not add up to the total semileptonic rate.
%abw Secondly, 
Second, 
$g^-$
depends on the relative abundance of 
various % four 
possible $D^{**}$ states, because
some of them decay only to $D^*\pi$ and others to $D\pi$,
depending on the spin and parity.
%abw The 
This relative %kroll
abundance is not measured very well at present.
Changing the abundance is equivalent to changing the
branching fractions 
%${\cal B}( D^{**} \rightarrow D^* \pi)$
%and ${\cal B}( D^{**} \rightarrow D \pi)$ 
for $D^* \pi$ and $D \pi$
averaged over various $D^{**}$ states. 
We define a quantity $P_V$ as
\[
P_V= \frac { {\cal B} (D^{**}\to D^*\pi) } 
    { {\cal B} (D^{**}\to D^* \pi) + {\cal B} (D^{**}\to D \pi)  } ,
\]
where ${\cal B}$ denotes a branching fraction.
We assume the relative abundance predicted %abw found 
in  Ref.~\cite{ISGW},
which corresponds to $P_V = 0.78$.
We also consider the extreme values $P_V = 0.0$ and $1.0$.
%Thirdly,
Third,
the composition depends on the 
ratio of the $B^-$ and ${\overline B}$$^0$ meson lifetimes,
because the number of $\ell^- D^{(*)}$ events 
is proportional to the semileptonic branching fraction, 
which is the product of the lifetime and 
the semileptonic partial width.
Finally, the sample composition
depends on 
%hans an experimental effect,
the reconstruction efficiency of the low energy pion
in the decay $D^{*+} \rightarrow D^0 \pi^+$. 
If we miss the pion and reconstruct the $D^0$ meson, 
the $D^{*+}$ decay is included in the $\ell^-D^0$ sample
and the sample  composition is altered.
The efficiency 
is measured %estimated 
to be $\epsilon(\pi) = 0.93 ^{+0.07} _{-0.21}$
by studying the rates of $\ell^- D^{*+}$ events
with respect to $\ell^- D^{0}$ events.

%%%% july 1997, plots of g- vs f** etc ????/ %%%%%%%%%%%%%%%%%%%%%%%%%%%

We also have to take into account the differences in the reconstruction
efficiencies for
the  ${\overline B} \rightarrow \ell^- {\bar \nu} D$, $D^*$ and $D^{**}$
decay modes. 
We examine this effect by using the Monte Carlo events
we have used to obtain the $K$ distributions. The ISGW model was used
for semileptonic decays.
We find that the first two modes show very similar efficiencies, 
while the last mode has an efficiency that is lower by  
about  a factor of two.
%% future %% 30 - 40\%, depending on the $D^{**}$ species and the lepton momentum.

\begin{figure}[p]
\vspace{1in}

\epsfysize=7.0in
\epsffile[ 50 144 522 648]{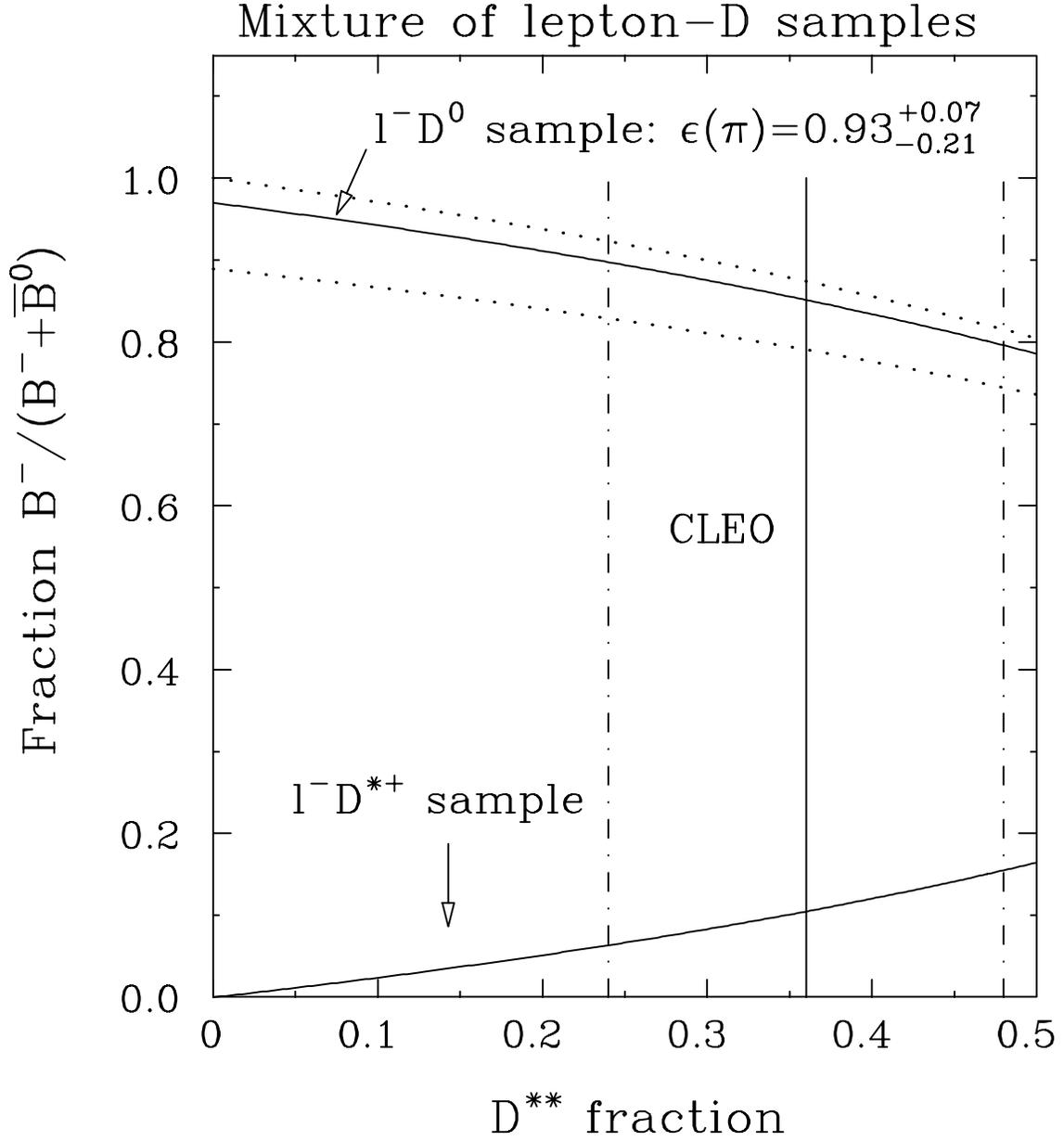}
\vspace{1in}
\caption 
{Fraction $g^-$ of $B^-$ mesons 
in lepton-$D^{(*)}$ samples 
as a function of the $D^{**}$ meson fraction $f^{**}$
in semileptonic $B$ decays.
Vertical lines show the range 
of CLEO measurement~{\protect \cite{CLEO}}.
The relative abundance of various $D^{**}$ mesons 
is fixed to $P_V =0.78$ (see text).
Low energy pion reconstruction efficiency 
is fixed to 0.93 (solid curves), 
0.72 and 1.0 (dotted curves).
}
\label {fig:frac_f2st}
\end {figure}

\begin{figure}[p]
\vspace{1in}
\epsfysize=7.0in
\epsffile[ 50 144 522 648]{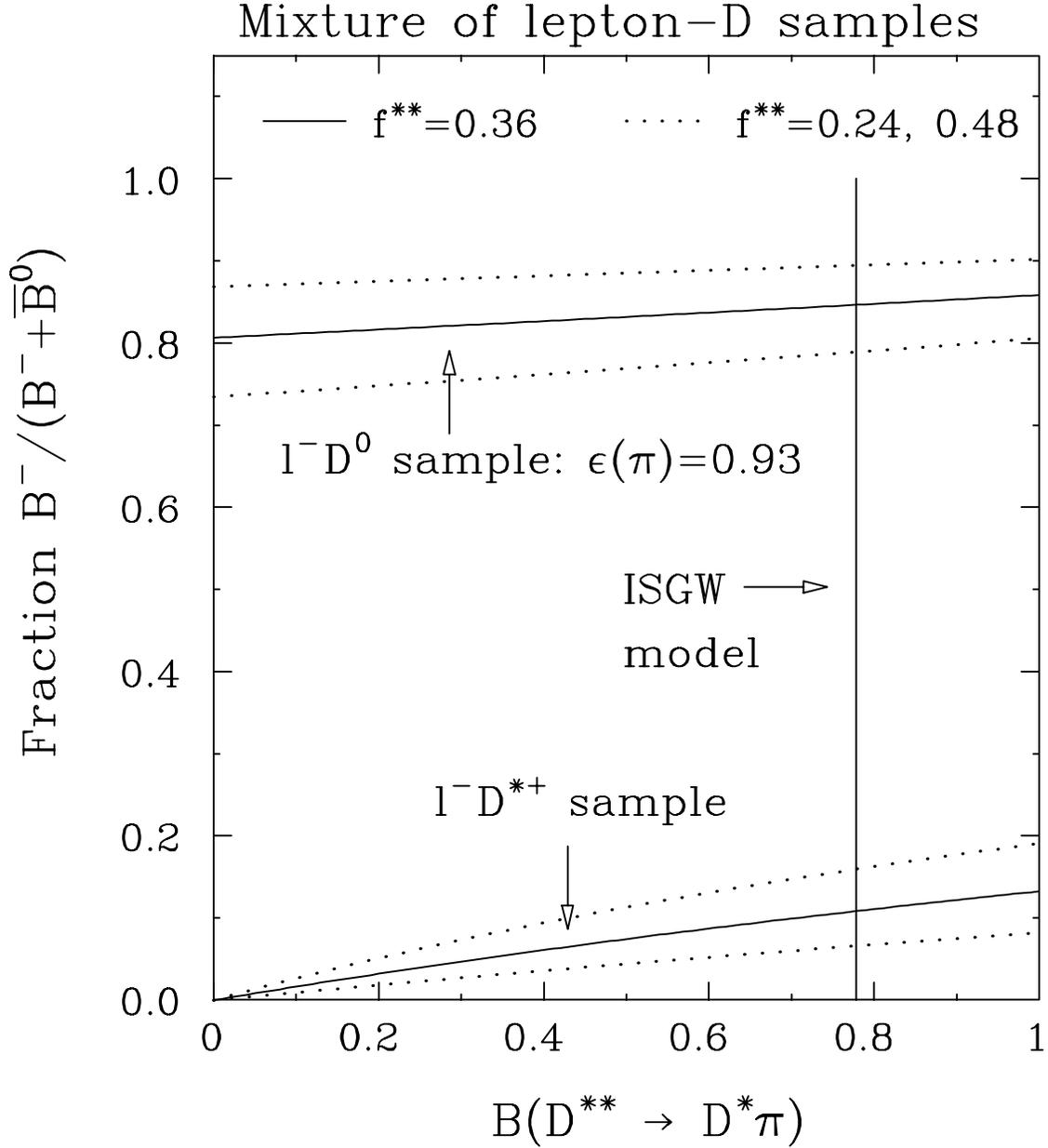}
\vspace{1in}
\caption 
{Fraction $g^-$ of $B^-$ mesons 
in lepton-$D^{(*)}$ samples 
as a function of the average $D^{**}$ branching fraction
${\cal B} ( D^{**} \rightarrow D^* \pi )$ or $P_V$.
Vertical line corresponds to the prediction of
the ISGW model~{\protect \cite{ISGW}}.
The $D^{**}$ fraction ($f^{**}$) 
is fixed to 0.36 (solid curves), 
0.24 and 0.48 (dotted curves).
Low energy pion reconstruction efficiency
is fixed to 0.93.}
\label {fig:frac_pv}
\end {figure}

The dependence of the $B^-$ fraction $g^-$ 
on the parameters $f^{**}$ and $P_V$ 
are illustrated in Figs.~\ref{fig:frac_f2st}
and \ref{fig:frac_pv}.
We find that
$g^- = 0.85 \, ^{+0.05} _{-0.12} $ for the $\ell^-D^0$ sample
and $g^- = 0.10 \, ^{+0.09} _{-0.10}$
for the $\ell^-D^{*+}$ sample 
when the two lifetimes are identical.
The central values correspond to the nominal choice
of the parameters, $f^{**} = 0.36$, $P_V = 0.78$
and $\epsilon(\pi) = 0.93$.
The %%% mar 20 98  quoted 
uncertainties reflect maximum changes in $g^-$
when 
%may 5 $f^{**}$, $P_V$ and $\epsilon(\pi)$ 
the three parameters
are changed 
within 
their %added may 5
%may 5 quoted 
%franco ranges.
uncertainties, namely $f^{**}$ to 0.24 and 0.48,
$P_V$ to 0.0 and 1.0, and $\epsilon(\pi)$ to 0.72 and 1.0.

We also note that 
the momentum correction factors ($K$ distributions)
need to be modified %%accordingly 
when the sample composition parameters
are changed. 
The $K$ distributions for the decay
${\overline B} \rightarrow \ell^- {\bar \nu} D^{**}$
%%% are softer %% Hans W
have lower mean values
because of additional missing particle(s), %% pion(s),
and changing the amount of $D^{**}$ decays  %%sample composition parameters
results in changes in the $K$ distributions.

There are other physics processes that can produce the lepton-$D^{(*)}$
signature. 
The largest background comes from the decay
of the ${\overline B}$$^0_s$ meson, 
${\overline B}$$^0_s  \rightarrow \ell^- {\overline \nu} D_s^{**+}$,
followed by $D_s^{**+} \rightarrow D^{(*)} K$.
%This process is estimated to contribute to about 2\% of the
%lepton-$D^{(*)}$ signal. 
The contribution of this process to the 
lepton-$D^{(*)}$ signal 
is estimated to be about 2\%.
Other processes such as %%may 98 like 
${\overline B} \rightarrow \tau^- {\bar \nu}_\tau D^{(*)} X$
followed by $\tau^- \rightarrow \ell^- {\bar \nu}_\ell \nu_{\tau}$,
and 
${\overline B} \rightarrow D_s^{-} D^{(*)} X$
followed by $D_s^{-} \rightarrow \ell^- X$,
are suppressed severely
because of 
branching fractions and
kinematic requirements on leptons.
We have ignored these backgrounds here.
Therefore the fraction of 
${\overline B}$$^0$ mesons is given by $g^0 = 1 - g^-$.
We treat effects of the physics backgrounds
as a systematic uncertainty.
%%%%%%%%%%% July 1997:  real D and a fake lepton????/?? %%%%%%%%%%

\subsection {Lifetime fit}

We can now determine the $B^-$ and ${\overline B}$$^0$ lifetimes
with a combined fit of the $\ell^-D^{0}$  and 
$\ell^-D^{*+}$  samples. The likelihood is given by
\begin{displaymath}
{\cal L} =  \prod_{\rm sample}  \left\{ 
\prod_i \, [ \, (1-f_{\rm BG}) {\cal F}_{\rm SIG}(x_i)
 + f_{\rm BG} {\cal F}_{\rm BG} (x_i) \, ]  \;
\prod_k {\cal F}_{\rm BG} (x_k) \right\},
\end{displaymath}
where the product is taken over event $i$ in each signal sample,
event $k$ in each background sample,
and over 
the $\ell^- D^0$ and $\ell^- D^{*+}$ samples.
For each signal sample,
we use a two-component signal distribution function given by
\[
{\cal F}_{\rm SIG}(x) = g^-  {\cal F}_{\rm SIG}^- (x)
                 + (1-g^-) {\cal F}_{\rm SIG}^0 (x),
\]
where 
${\cal F}_{\rm SIG}^- (x)$ and ${\cal F}_{\rm SIG}^0 (x)$ represent
the $B^-$ and ${\overline B}$$^0$ meson components, respectively.
%%The fraction $g^-$ of the $B^-$ meson decays 
%%in each sample is 
%%% very different for the two lepton-$D$ samples,
The dependence of $g^-$
on the lifetime ratio
is taken into account during lifetime fits.

The result of the combined fit is
$c\tau (  B^- ) = 
491 \pm 17$ $\mu$m, 
$c\tau ( {\overline B}$$^0) = 
%444 \pm 12$ $\mu$m, 
442 \pm 12$ $\mu$m, 
where the quoted uncertainties are statistical, and are
correlated with each other with a coefficient of 
$-0.308$.
From these numbers we calculate the ratio of the lifetimes to be
$\tau ( B^- ) / \tau ( {\overline B}$$^0) = 
%1.106 \pm 0.055$. 
1.110 \pm 0.056$. 

\begin{figure}[p]
\vspace{1in}
\epsfysize=7.0in
\epsffile[ 50 144 522 648]{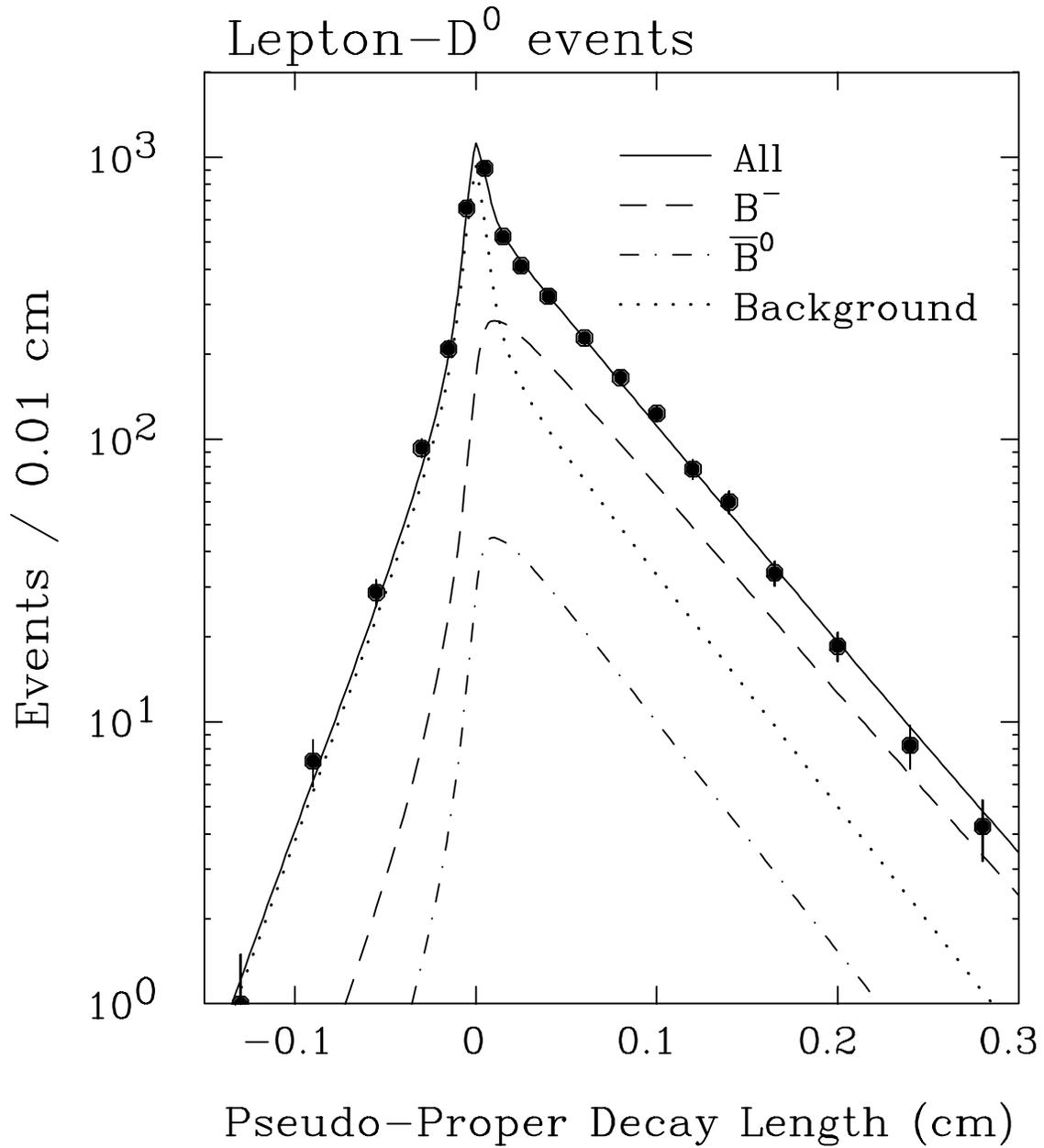}
\vspace{1in}
\caption{ 
Pseudo-proper decay length distribution 
of the $\ell^- D^0$ candidates (points).
Curves show the result of the combined fit
with $\ell^- D^{*+}$ candidates:
The $B^-$ component (dashed curve),
the ${\overline B}$$^0$ component (dot-dashed curve),
and the background component (dotted curve).
}
\label {fig:life_bpl_b0_d0}
\end {figure}

\begin{figure}[p]
\vspace{1in}
\epsfysize=7.0in
\epsffile[ 50 144 522 648]{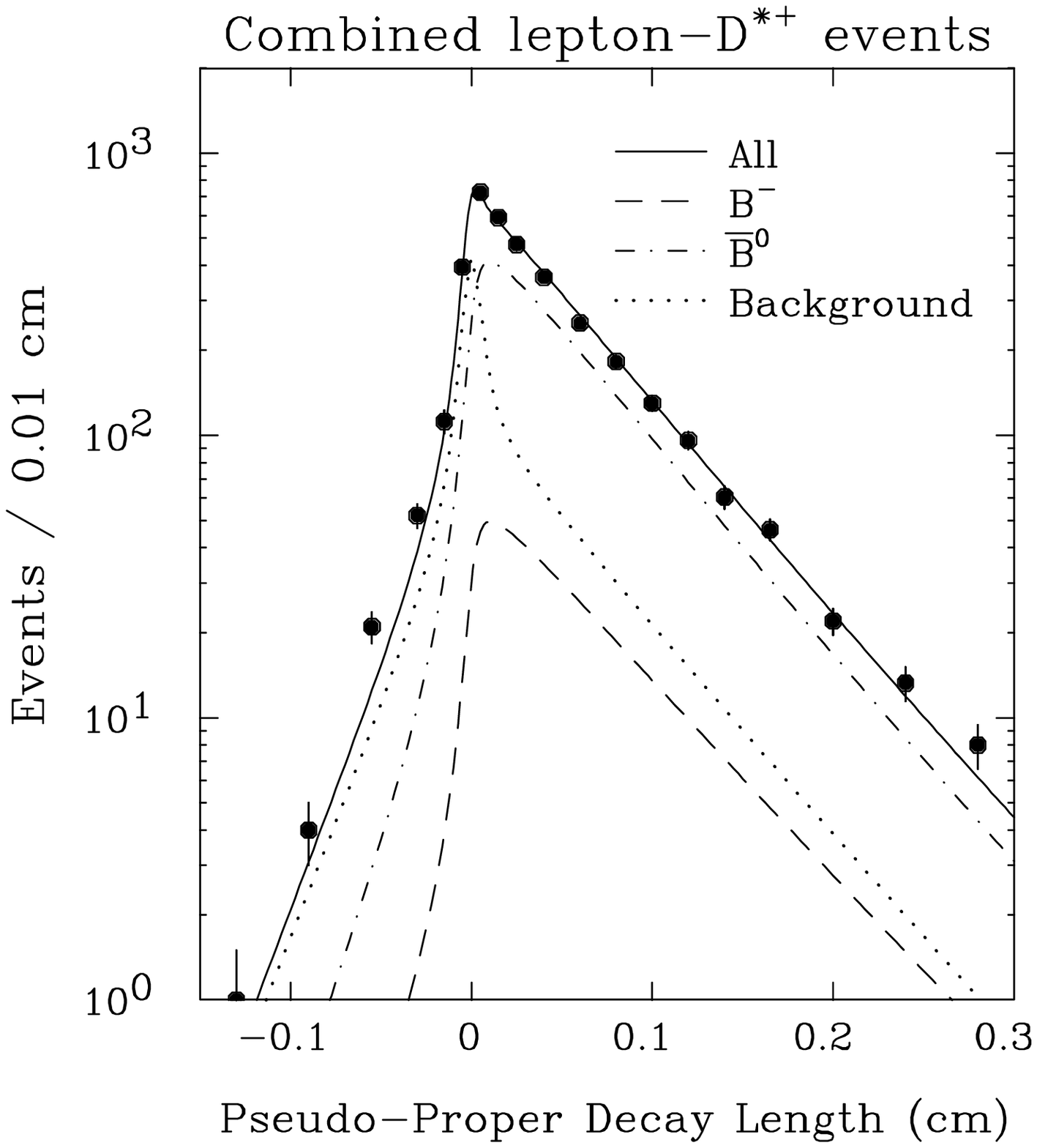}
\vspace{1in}
\caption{ 
Pseudo-proper decay length distribution 
of the $\ell^- D^{*+}$ candidates (points).
The three $D^0$ decay modes are combined.
Curves show the result of the combined fit
with $\ell^- D^{0}$ candidates:
The ${\overline B}$$^0$ component (dot-dashed curve),
the $B^-$ component (dashed curve),
and the background component (dotted curve).
}
\label {fig:life_bpl_b0_dst}
\end {figure}

The pseudo-proper decay length distributions of the $\ell^- D^0$ sample
and the combined $\ell^- D^{*+}$ sample are shown 
in Figs.~\ref{fig:life_bpl_b0_d0}
and \ref{fig:life_bpl_b0_dst}.
The results of the combined fit are superimposed.

\subsection {Systematic uncertainties}

The sample composition is a source
of systematic uncertainty in the $B$ meson lifetime determination.
We change 
% for harvard may 5 98 %++
each one of
the parameters $f^{**}$, $P_V$ and 
$\epsilon(\pi)$ 
to another value 
while keeping others
at their nominal values,
% within the quoted ranges,
compute the sample composition $g^-$ 
and fit for the two $B$ meson lifetimes.
The results are listed in Table~\ref{tb:sample}.
We interpret the observed changes as systematic uncertainties.

Other sources of systematic uncertainties 
considered in this analysis are described below.
They are summarized in Table~\ref{tb:systematics}.

%hans, remove
%The amount of the background in the signal sample,
%as well as the shapes of the background decay length distributions,
%are subject to uncertainty because they are determined with
%finite statistical precision.
%We have included them as fitting parameters 
%%%(although constrained)
%in the lifetime fits, and therefore their effects are
%absorbed in statistical uncertainties in the lifetimes.
%The way the background decay length distributions
%are modeled %%described
%is also a source of a systematic uncertainty.
%

We have estimated the decay length distributions
from real data using mass sidebands, 
thus minimizing model dependence.
However, the assumed functional form
may not be fully adequate to describe the true shapes.
Thus, we have considered an alternative parameterization
that includes additional exponential terms;
this has turned out to give only minimal changes in the result.

Physics and fake lepton background processes
are studied by adding their simulated decay
length distributions to the background function.
%% mar 98
%The largest contribution
%comes from the decay
%${\overline B}$$^0_s  \rightarrow \ell^- {\overline \nu} D_s^{**+}$,
%followed by $D_s^{**+} \rightarrow D^{(*)} K$,
%and accounts for about 2\% of observed $\ell^- D^{(*)}$ signals.
%Other processes like 
%${\overline B} \rightarrow \tau^- {\bar \nu}_\tau D^{(*)} X$
%followed by $\tau^- \rightarrow \ell^- {\bar \nu}_\ell \nu_{\tau}$
%and 
%${\overline B} \rightarrow D_s^{-} D^{(*)} X$
%followed by $D_s^{-} \rightarrow \ell^- X$
%are suppressed severely
%because of 
%branching fractions and
%kinematic requirements on leptons.

Other sources of systematic uncertainties 
include our estimate of
the decay length resolution 
and of the $B$ meson momentum.
We have introduced a resolution scale factor $s$ and find a value
of about 1.35. 
We change this factor to 1.0 or 1.7, fix it at the value
and repeat the lifetime fitting procedure. 
%We find the changes of ?? (??) $\mu$m in the $B^-$ (${\overline B}$$^0$) meson
%lifetime. 
We assign the observed changes as an uncertainty.
The momentum correction ($K$ distribution) is
subject to some uncertainty too, 
because it depends on
the kinematics of $B$ meson production 
and %%% on the model 
of semileptonic decays.
%harv may 5 We investigate different production and decay models 
An alternative $p_T$ spectral shape of the $b$ quark production
was considered, based on a comparison of lepton $p_T$ shape 
in the real data and Monte Carlo events.
A simple $V-A$ decay model was tried in place of the ISGW model
to describe semileptonic decays.
We apply these changes and obtain new
$K$ distributions, and repeat the lifetime fits.
The observed changes are listed as a systematic uncertainty.
In addition, the $K$ distributions 
are somewhat dependent on the lepton momentum
and on the cuts used for electron identification.
We assign uncertainties due to possible incompleteness 
in the treatment of these effects.
As stated earlier, 
the momentum correction depends on
the assumed amount of $B$ decays to $D^{**}$ mesons.
This effect is already accounted for in the sample
composition uncertainty.

Also, 
we have applied a loose cut 
on the $D^0$ decay length  %, may 5
in some modes, %when we reconstruct some of the charm modes,
and it introduces a slight bias 
%(about 5 $\mu$m) 
(about 2.5 $\mu$m) 
%fumi Feb 98in the lifetimes.
toward a longer lifetime.
Here we quote the number without correction to the final lifetimes
%fumi Feb 98
and assign a systematic uncertainty.
Finally, a possible residual misalignment of the SVX detector
and the stability of the position of the Tevatron beam
are considered.
%as in other CDF lifetime measurements.
Some of these uncertainties 
are common to
the two $B$ mesons and 
cancel in the determination of the lifetime ratio.
All these effects are combined  in quadrature to give
the total systematic uncertainty.

\section {Final results and conclusion}
We have measured the lifetimes of the
$B^-$ and ${\overline B}$$^0$ mesons 
using their
partially reconstructed semileptonic decays
$\overline{B} \rightarrow \ell^- {\overline \nu} D^0 X$  and 
$\overline{B} \rightarrow \ell^- {\overline \nu} D^{*+} X$.
Our final results are
\begin{eqnarray*}
%\tau(B^-)                & = & 1.637 \pm 0.057 \pm 0.053 \;\; {\rm ps}, 
%\tau(B^-)                & = & 1.637 \pm 0.057 \pm 0.051 \;\; {\rm ps}, 
%\tau(B^-)                & = & 1.637 \pm 0.058 \pm 0.051 \;\; {\rm ps}, 
\tau(B^-)                 & = & 1.637 \pm 0.058 \, ^{+\, 0.045} _{ -\, 0.043} \;\; {\rm ps},
							\nonumber \\
%\tau({\overline B}\,\!^0) & = & 1.480 \pm 0.040 \pm 0.054 \;\; {\rm ps}, 
%\tau({\overline B}\,\!^0) & = & 1.480 \pm 0.040 \pm 0.057 \;\; {\rm ps}, 
%\tau({\overline B}\,\!^0) & = & 1.474 \pm 0.039 \pm 0.057 \;\; {\rm ps}, 
 \tau({\overline B}\,\!^0) & = & 1.474 \pm 0.039 \, ^{+\, 0.052} _{ -\, 0.051} \;\; {\rm ps},
							\nonumber \\
%\tau(B^-) / \tau({\overline B}\,\!^0) & = & 1.106 \pm 0.055  
\tau(B^-)  / \tau({\overline B}\,\!^0) & = & 1.110 \pm 0.056
%\, ^{+\, 0.031} _{ -\, 0.028} ,
%\, ^{+\, 0.033} _{ -\, 0.029} ,
\, ^{+\, 0.033} _{ -\, 0.030} ,  % may 6, 98
\end{eqnarray*}
where 
the first uncertainties are statistical and the second are systematic.
%%% Andy This measurement supersedes a previous CDF measurement~\cite{semi_1a}.
The result is consistent with
other recent measurements~\cite{bu_bd_life,cdf_psi_new}.
We combine this measurement with the CDF measurement~\cite{cdf_psi_new}
using fully reconstructed decays,
\begin{eqnarray*}
\tau(B^-)                 & = & 1.68 \pm 0.07 \pm 0.02 \;\; {\rm ps}, 
							\nonumber \\
\tau({\overline B}\,\!^0)  & = & 1.58 \pm 0.09 \pm 0.02 \;\; {\rm ps}, 
							\nonumber \\
\tau(B^-)  / \tau({\overline B}\,\!^0) & = & 1.06 \pm 0.07 \pm 0.02,
\end{eqnarray*}
and derive the following CDF  average:
\begin{eqnarray*}
%\tau(B^-)                 & = & 1.663 \pm 0.053 \;\; {\rm ps}, 
%\tau(B^-)                 & = & 1.661 \pm 0.053 \;\; {\rm ps}, 
%\tau(B^-)                 & = & 1.662 \pm 0.053 \;\; {\rm ps},  %may 6 98
 \tau(B^-)                 & = & 1.661 \pm 0.052 \;\; {\rm ps},  %Jun 8 98
							\nonumber \\
%\tau({\overline B}\,\!^0) & = & 1.520 \pm 0.057 \;\; {\rm ps}, 
%\tau({\overline B}\,\!^0) & = & 1.517 \pm 0.055 \;\; {\rm ps}, 
%\tau({\overline B}\,\!^0) & = & 1.517 \pm 0.056 \;\; {\rm ps}, %may 98
 \tau({\overline B}\,\!^0) & = & 1.513 \pm 0.053 \;\; {\rm ps}, %jun 98
							\nonumber \\
%\tau(B^-) / \tau({\overline B}\,\!^0) & = & 1.089 \pm 0.049,
%\tau(B^-) / \tau({\overline B}\,\!^0) & = & 1.088 \pm 0.049,
%\tau(B^-) / \tau({\overline B}\,\!^0) & = & 1.090 \pm 0.050,
 \tau(B^-) / \tau({\overline B}\,\!^0) & = & 1.091 \pm 0.050,
\end{eqnarray*}
where the uncertainties include both statistical and systematic
effects.
There exists a small (about 3 $\mu$m) 
correlation in systematic effects
between the two measurements,
such as due to detector alignment,
and it is taken into account in combining the results.

%abw At present, 
The ratio of the two $B$ meson lifetimes
differs from unity by about 
9\%, or % this added Mar 98
two standard deviations.
%franco    but the difference is not significant yet.
%% fumi Feb 98
This 
agrees with the small difference predicted by theory.
The result is also
consistent with the current world average value
of $1.03  \pm 0.05$~\cite{PDG}.
The ${\overline B}$$^0$   meson lifetime 
is consistent with  %hans identical to 
the ${\overline B}$$^0_s$ meson lifetime~\cite{bs_life}
within the uncertainty.

\begin{acknowledgements}
     We thank the Fermilab staff and the technical staffs of the
participating institutions for their vital contributions. This work was
supported by the U.S. Department of Energy and National Science 
Foundation;
the Italian Istituto Nazionale di Fisica Nucleare;
the Ministry of Education, Science and Culture of Japan;
the Natural Sciences and Engineering Research Council of Canada;
the National Science Council of the Republic of China;
the A. P. Sloan Foundation;
the Swiss National Science Foundation.
%%and the German Bundesministerium f\"{u}r Bildung, Wissenschaft, 
%%Forschung, und Technologie.
%%%and the Alexander von Humboldt-Stiftung.
\end{acknowledgements}

%%\vspace{0.3in}
\renewcommand{\baselinestretch}{1.00}
%%%  References
%%%\begin{references}
%%{\noindent $^{a}$ Visitor}

\rm

\clearpage %%%%%%%%%%%%%% ttables

%\begin {table}
%\begin{center}
%\begin {tabular} {lccccc}
% Mode & 
%\multicolumn{2}{c} {cone size} & \multicolumn{2}{c} {momentum (GeV/$c$) }
%& Decay \\ %& \multicolumn{2}{c} {Isolation }  & Vertex $\chi^2$ \\
% & kaon & pion & kaon & pion 
%& length \\ \hline %& $e$ & $\mu$  & \\ \hline

%$\ell^-D^0, D^0 \rightarrow K^- \pi^+ $ 
%&    0.6 & 0.7    & $> 1.5$  & $> 0.5$  & $> 0$ \\

%$\ell^- D^{*+}, D^0 \rightarrow  K^- \pi^+$ 
%&    0.6 & 0.7    & $> 1.5$  & $>0.5$  & - \\

% $\ell^- D^{*+}, D^0 \rightarrow K^- \pi^+\pi^+\pi^-$
%&   0.65 & 1.0    & $>1.2$  & $>0.5$  & $> 0$ \\

%$\ell^- D^{*+}, D^0 \rightarrow K^- \pi^+ \pi^0$
%&   0.6 & 0.7    & $>1.5$  & $>0.5$  & $> 0$ \\
%\end{tabular}
%\end{center}
%\caption { A summary of cuts used for charm meson reconstruction.
%}
%\label {tb:charm_cuts}
%\end{table}
%

\begin {table}
\begin{center}
\begin {tabular} {llccrc} 
 $B$ Mode & $D^0$ mode		& $D^0$ mass range & $\Delta m$ range 
& Events %Candidates
& Background fraction \\
&			& (GeV/$c^2$)      & (GeV/$c^2$)
& & \\ \hline
$\ell^-D^0$ %, D^0 \rightarrow 
& $K^- \pi^+ $ 
&  1.84 $-$ 1.88   &  Not $D^{*+}$
& 5198 & $0.526 \pm 0.018$ \\

$\ell^- D^{*+}$ %, D^0 \rightarrow  
& $K^- \pi^+$ 
&   1.83 $-$ 1.90 & 0.144 $-$ 0.147
& 935 & $0.086 \pm 0.011$ \\

 $\ell^- D^{*+} $%, D^0 \rightarrow %
& $K^- \pi^+\pi^+\pi^-$
&    1.84 $-$ 1.88 & 0.144 $-$ 0.147
& 1166 	& $0.183 \pm 0.015$ \\ 

$\ell^- D^{*+} $% , D^0 \rightarrow 
& $K^- \pi^+ \pi^0$
&    1.50 $-$ 1.70  & $< 0.155$
%& 2820 & $0.366 \pm 0.016$ \\ 
& 2858 & $0.366 \pm 0.016$ \\ 
\end{tabular}
\end{center}
\caption { Definition of signal samples, 
numbers of candidates and estimated background fraction.
}
\label {tb:signal}
\end{table}

\begin {table}
\begin{center}
\begin {tabular} {llccr} 
$B$ Mode & $D^0$ mode & $D^0$ mass range 
%& \multicolumn{2}{c} { $\Delta m$ range {(GeV/$c^2$)} }
& $\Delta m$ range 
& Events  \\ %\cline{4-5}
  &    & (GeV/$c^2$)      
%& $D^0 \pi^+$ & $D^0 \pi^-$  & \\ \hline
%& $D^0 \pi^+$ & $D^0 \pi^-$  
& (GeV/$c^2$)
& \\ \hline

$\ell^-D^0$%, D^0 \rightarrow 
&$K^- \pi^+ $ 
%&     1.74 $-$ 1.79,  1.94 $-$ 1.99  
%%&     1.72 $-$ 1.80, 1.92 $-$ 2.00
 &     1.74 $-$ 1.79,  1.94 $-$ 1.99  
%&   \multicolumn{2}{c} { Not $D^{*+}$ } &  7200 \\
&   Not $D^{*+}$ &  7200 \\

$\ell^- D^{*+}$%, D^0 \rightarrow  
&$K^- \pi^+$ 
&   1.83 $-$ 1.90  &  0.15 $-$ 0.19  %& $<$ 0.19  
%& 3377 \\
& 1769 \\

 $\ell^- D^{*+}$%, D^0 \rightarrow 
&$K^- \pi^+\pi^+\pi^-$
&   1.84 $-$ 1.88  &  0.15 $-$ 0.19 %&   $<$ 0.19  
%& 9544 \\
& 5030 \\

$\ell^- D^{*+}$%, D^0 \rightarrow 
&$K^- \pi^+ \pi^0$
&   1.50 $-$ 1.70  &  0.16 $-$ 0.19  %&   $<$ 0.19  
%& 7900 \\
& 3809 \\
\end{tabular}
\end{center}
\caption { Definition of background samples
and numbers of events.
}
\label {tb:control}
\end{table}

%%%%%% 
\begin {table}
\begin{center}
\begin {tabular} {llccccc}  %\hline
 $B$ Mode & $D^0$ mode & scale $s$ & $ f_+$ & $\lambda_+$ ($\mu$m)  
			& $ f_-$ & $\lambda_-$ ($\mu$m)  %& $f_{\rm BG}$
\\ \hline 
%\multicolumn{2}{l}{$\ell^-D^0, D^0 \rightarrow K^- \pi^+ $} & & & & & \\ 

$\ell^-D^0$%, D^0 \rightarrow 
&$K^- \pi^+ $
& $1.38 \pm 0.03$ & $0.404 \pm 0.008$ & $ 531 \pm 12$
			 & $0.136 \pm 0.007$ & $ 240 \pm 10$\\
%& $0.518 \pm 0.012$ \\ 

$\ell^- D^{*+}$%, D^0 \rightarrow  
&$K^- \pi^+$ 
% & $1.37 \pm 0.05$ & $0.487 \pm 0.012$ & $ 508 \pm 15$
%			 & $0.127 \pm 0.010$ & $ 273 \pm 20$ \\
 & $1.32 \pm 0.07$ & $0.487 \pm 0.017$ & $ 498 \pm 21$
			 & $0.136 \pm 0.014$ & $ 240 \pm 22$ \\

 $\ell^- D^{*+}$%, D^0 \rightarrow 
&$K^- \pi^+\pi^+\pi^-$
%& $1.36 \pm 0.02$ & $0.302 \pm 0.008$ & $ 372 \pm \:\:9$
%			  & $0.051 \pm 0.005$ & $ 236 \pm 18$\\
& $1.38 \pm 0.03$ & $0.328 \pm 0.011$ & $ 362 \pm 12$
			  & $0.058 \pm 0.008$ & $ 216 \pm 21$\\

$\ell^- D^{*+}$%, D^0 \rightarrow 
&$K^- \pi^+ \pi^0$
%& $1.37 \pm 0.03$ & $0.526 \pm 0.008$ & $ 599 \pm 12$ 
%			 & $0.107 \pm 0.006$ & $ 258 \pm 13$ \\
& $1.39 \pm 0.04$ & $0.536 \pm 0.011$ & $ 612 \pm 17$ 
			 & $0.098 \pm 0.008$ & $ 274 \pm 20$ \\
\end{tabular}
\end{center}
\caption { Background shapes obtained from a simultaneous fit
of signal and background samples. 
}
\label {tb:shapes}
\end{table}

%%%%%% 
\begin {table}
\begin{center}
\begin {tabular} {llcc}  %\hline
$B$ Mode & $D^0$ mode & $c\tau(B)$ ($\mu$m) & $c\tau(D^0)$ ($\mu$m) 
\\ \hline 

$\ell^-D^0$%, D^0 \rightarrow 
&$K^- \pi^+ $
& $489 \pm 15$ 
%& $124.3 \pm 5.6$ \\
 & $128.0 \pm 5.3$ \\

$\ell^- D^{*+}$%, D^0 \rightarrow  
&$K^- \pi^+$ 
%& $461 \pm 18$ 
& $462 \pm 18$ 
%& $132.9 \pm 5.7$ \\
% & $132.5 \pm 5.6$ \\
 & $133.8 \pm 5.6$ \\  % mar 98 RS BG only

 $\ell^- D^{*+}$%, D^0 \rightarrow 
&$K^- \pi^+\pi^+\pi^-$
%& $472 \pm 19$ 
& $472 \pm 19$ 
%& $126.0 \pm 5.4$ \\
% & $126.3 \pm 5.2$ \\
 & $125.3 \pm 5.2$ \\  % mar 98

$\ell^- D^{*+}$%, D^0 \rightarrow 
&$K^- \pi^+ \pi^0$
%& $453 \pm 15$ 
& $449 \pm 14$ 
%& $120.9 \pm 5.7$ \\
% & $125.4 \pm 5.1$ \\
 & $127.5 \pm 5.0$ \\
\end{tabular}
\end{center}
\caption { $B$ and $D^0$ meson lifetimes 
%obtained 
measured
% from a simultaneous fit of signal and background samples 
for individual decay modes.
Quoted uncertainties are statistical only.
}
\label {tb:lifetimes}
\end{table}

\begin {table}
\begin{center}
\begin {tabular} {ccccccccc} 
$f^{**}$  & $P_V$ & $\epsilon( \pi )$ 
& \multicolumn{2}{c} {$g^-$ } 
& \multicolumn{2}{c} {$c \tau$  ($\mu$m) } 
& correl. & $\tau ( B^- )$ \\ \cline{6-7}

& & &$\ell^- D^0$ & $\ell^- D^{*+}$ &
 $B^-$ & ${\overline B}$$^0$  & coeff. 
& ${ \overline { \tau ( {\overline B}\,\!^0 ) } } $ \\ \hline 

0.24 & 0.78 & 0.93 & 0.899 & 0.064 & 
%$491.2 \pm 16.3 $ & $449.6 \pm 10.6 $ & $-0.187$ & $1.093 \pm 0.048$ \\ 
 $491.3 \pm 16.3 $ & $448.0 \pm 10.7 $ & $-0.187$ & $1.097 \pm 0.049$ \\ 

0.36 & 0.78 & 0.93 & 0.851 & 0.105 & 
%$490.8 \pm 17.3$ & $443.9 \pm 11.5 $ & $-0.308$ & $1.106 \pm 0.055$\\
 $491.0 \pm 17.3$ & $442.2 \pm 11.6 $ & $-0.308$ & $1.110 \pm 0.056$\\

0.48 & 0.78 & 0.93 & 0.796 & 0.155 & 
%$491.6 \pm 18.9$ & $436.1 \pm 13.1 $ & $-0.461$ & $1.127 \pm 0.066$ \\ \hline
 $492.0 \pm 18.9$ & $434.2 \pm 13.3 $ & $-0.461$ & $1.133 \pm 0.067$ \\ \hline

0.36 & 0.00 & 0.93 & 0.806 & 0.000 & 
%$490.9 \pm 17.5$ & $449.6 \pm \:\:9.6 $ & $-0.105$ & $1.092 \pm 0.048$ \\
 $491.0 \pm 17.5$ & $448.2 \pm \:\:9.7 $ & $-0.105$ & $1.096 \pm 0.048$ \\

0.36 & 1.00 & 0.93 & 0.858 & 0.133 & 
%$491.0 \pm 17.3$ & $442.0 \pm 12.2 $& $-0.360$ & $1.111 \pm 0.058$ \\  \hline
 $491.3 \pm 17.4$ & $440.2 \pm 12.3 $& $-0.360$ & $1.116 \pm 0.058$ \\  \hline

0.36 & 0.78 & 0.72 & 0.790 & 0.105 & 
%$494.2 \pm 18.7$ & $443.4 \pm 11.6 $& $-0.357$ & $1.114 \pm 0.059$ \\ 
 $494.5 \pm 18.7$ & $441.7 \pm 11.8 $& $-0.357$ & $1.120 \pm 0.060$ \\ 

0.36 & 0.78 & 1.00 & 0.874 & 0.105 & 
%$489.7 \pm 16.8$ & $444.0 \pm 11.4 $ & $-0.290$ & $1.103 \pm 0.054$ \\  
 $489.8 \pm 16.8$ & $442.3 \pm 11.5 $ & $-0.290$ & $1.107 \pm 0.054$ \\  
\end{tabular}
\end{center}
\caption{
$B^-$ and ${\overline B}$$^0$ lifetimes 
from a combined fit 
of $\ell^- D^0$ and $\ell^- D^{*+}$ samples 
under various sample composition conditions.
Quoted uncertainties are statistical only 
and are correlated between $B^-$ and ${\overline B}$$^0$.
Also listed are their calculated ratios.
}
\label {tb:sample}
\end{table}

\begin{table}[ht]
\begin{center}
\begin {tabular} {lccc} %%%% \hline 
% Source & \multicolumn{3}{c} {Systematic uncertainty contributions to} \\ 
 Source & \multicolumn{3}{c} {Contribution to} \\ 
%%&  $c\tau( B^-)$ ($\mu$m)  &  $c\tau( {\overline B}$$^0)$ ($\mu$m)  
%%&  $\tau( B^-) / \tau( {\overline B}^0 ) $ \\  
&  $c\tau( B^-)$ &  $c\tau( {\overline B}$$^0)$ &  $\tau( B^-) $ \\ 
% next line
 & ($\mu$m)  & ($\mu$m)  & ${\overline  {\tau( {\overline B}\,\!^0 ) }}$ \\
\hline

 Sample composition & & \\
 \hspace{4mm} $D^{**}$ fraction ($f^{**}$)
%may 7  & $ ^{+1} _{-0} $  &   $ ^{+6}_{-8} $  & $ ^{+0.022} _{-0.013}$\\
        & $ ^{+1} _{-0} $  &   $ ^{+6}_{-8} $  & $ ^{+0.023} _{-0.014}$\\
 \hspace{4mm} $D^{**}$ composition ($P_V$)
%may 98 &  $ ^{+1} _{-0} $  &  $^{+6} _{-2} $   & $ ^{+0.005} _{-0.014}$ \\
        &  $ ^{+1} _{-0} $  &  $^{+6} _{-2} $   & $ ^{+0.006} _{-0.015}$ \\
 \hspace{4mm} Low energy pion reconstruction
%may 7 98 & $  ^{+4} _{-2} $  &  $^{+1} _{-2} $   & $ ^{+0.009} _{-0.003}$ \\
          & $  ^{+4} _{-1} $  &  $\pm 1$          & $ ^{+0.009} _{-0.003}$ \\

%  Background treatment  & $ \pm 2 $  & $ \pm 2 $         & $\pm 0.006$ \\
%% Apr 96, now includes 2 micron physics BG uncertainty (Bs -> l nu D**s)
%%   Background treatment  & $ \pm 3 $  & $ \pm 3 $         & $\pm 0.009$ \\
%% Feb 98, Add 3.7 microns due to fake lepton + real D, tot = 4.7, quote 5.
   Background treatment  & $ \pm 5 $  & $ \pm 5 $         & $\pm 0.015$ \\

%  Background shape     & $^{+9} _{-8}$  & $ \pm 6 $         & $\pm 0.024$ \\

%  Background fraction  & $ \pm 10 $    & $ \pm 11 $         & $\pm 0.033$ \\

% Physics background      &  $\pm  2$ & & \\

% Decay length resolution    & $^{+9} _{-8}$   & $^{+6} _{-4}$  
%& $^{+0.007} _{-0.009}$  \\
% This number hads been completely wrong. It was not even the final Ia number.
% It was a number from CDF 3009.
% 
 Decay length resolution    & $^{+7} _{-5}$   & $^{+7} _{-5}$  
& $ \pm 0.002 $  \\

% Momentum estimate           & $\pm 12$ & $\pm 12$    & - \\
 Momentum estimate           & & & \\
 \hspace{4mm} $b$ quark $p_T$ spectrum    &  $\pm  4$ &  $\pm  4$ &  -\\
 \hspace{4mm} $B$ decay model         &  $\pm  4$ &  $\pm  4$ & -\\
 \hspace{4mm} Momentum dependence     &  $\pm  6$ &  $\pm  6$ & -\\
%\hspace{4mm} Electron cuts           &  $\pm  9$ &  $\pm  9$ & -\\
 \hspace{4mm} Electron cuts           &  $\pm  5$ &  $\pm  5$ & -\\ % june 98
%order changed may 5 98
% \hspace{4mm} $B$ decay model         &  $\pm  4$ &  $\pm  4$ & -\\

Decay length cut         &  $^{+0} _{-5}$   & $^{+0} _{-5}$  & $\pm 0.016$ \\

 Detector alignment       &  $\pm   2$ &  $\pm   2$ &  - \\ \hline
% Beam stability           &  $\pm   5$ &  $\pm   5$ &  - \\ \hline

%feb98 Total    & $ \pm 16 $ &  $ \pm 16 $         & $^{+0.031} _{-0.028}$ \\
% Total    & $ \pm 15 $ &  $ \pm 17 $         & $^{+0.033} _{-0.029}$ \\
% Total    & $ \pm 15 $ &  $ \pm 17 $         & $^{+0.033} _{-0.030}$ \\
  Total    & $ \pm 13 $ &  $ ^{+16} _{-15} $  & $^{+0.033} _{-0.030}$ \\ % june 98
%%% prd \hline
\end{tabular}
\end{center}
\caption { A summary of systematic uncertainties in the
$B^-$ and ${\overline B}$$^0$ lifetime measurement.
}
\label {tb:systematics}
\end{table}

\clearpage
%%%%

%%%%% Schematic diagram
\clearpage

%%%%% k dist
\clearpage

%%%%%%%%%%%%%%%% bg decay length dist
\clearpage

%%%%%%%%%%%%%%%% sig decay length dist
\clearpage

%%%%%%%%%%%%%%%% sig decay length dist
\clearpage

%%%%%%%%%  sample composition figures
\clearpage

\clearpage

\clearpage

\clearpage


\begin{thebibliography}{99}
\bibitem{foot2} In CDF, $\varphi$ is the azimuthal angle, 
$\theta$ is the polar angle measured from the proton direction, 
and $r$ is the radius from the beam axis ($z$-axis).
\end{thebibliography}
\end{document}